\documentclass[a4paper,fleqn,usenatbib]{mnras}

\usepackage{newtxtext,newtxmath}
\usepackage[T1]{fontenc}
\usepackage{ae,aecompl}

\usepackage{graphicx}	
\usepackage{amsmath}	

\DeclareMathOperator*{\argmin}{arg\,min}

\title[Effect of emission lines on photo-z]{The effect of emission lines on the performance of photometric redshift estimation algorithms}

\author[G. Cs\"ornyei, L. Dobos \& I. Csabai]{
G\'eza Cs\"ornyei,$^{1,2,3}$\thanks{E-mail: csogeza@mpa-garching.mpg.de}
L\'aszl\'o Dobos,$^{1,4}$ 
Istv\'an Csabai$^{1}$\\
$^{1}$Department of Physics of Complex Systems, ELTE E\"otv\"os Lor\'and University, P\'azm\'any P\'eter s\'et\'any 1/a, Budapest 1117, Hungary \\
$^{2}$Konkoly Observatory, CSFK, Konkoly-Thege M. \'ut 15-17, Budapest, 1121, Hungary \\
$^{3}$Max Planck Institute for Astrophysics, Karl-Schwarzschild-Str. 1, 85741 Garching, Germany\\
$^{4}$Department of Physics and Astronomy, Johns Hopkins University, 3400 N. Charles Street, Baltimore, MD 21218, USA
}

\date{Accepted XXX. Received YYY; in original form ZZZ}

\pubyear{2021}

\begin{document}
\label{firstpage}
\pagerange{\pageref{firstpage}--\pageref{lastpage}}
\maketitle

\begin{abstract}
We investigate the effect of strong emission line galaxies on the performance of empirical photometric redshift estimation methods. In order to artificially control the contribution of photometric error and emission lines to total flux, we develop a PCA-based stochastic mock catalogue generation technique that allows for generating infinite signal-to-noise ratio model spectra with realistic emission lines on top of theoretical stellar continua. Instead of running the computationally expensive stellar population synthesis and nebular emission codes, our algorithm generates realistic spectra with a statistical approach, and -- as an alternative to attempting to constrain the priors on input model parameters -- works by matching output observational parameters. Hence, it can be used to match the luminosity, colour, emission line and photometric error distribution of any photometric sample with sufficient flux-calibrated spectroscopic follow-up. 
We test three simple empirical photometric estimation methods and compare the results with and without photometric noise and strong emission lines. While photometric noise clearly dominates the uncertainty of photometric redshift estimates, the key findings are that emission lines play a significant role in resolving colour space degeneracies and good spectroscopic coverage of the entire colour space is necessary to achieve good results with empirical photo-z methods.
Template fitting methods, on the other hand, must use a template set with sufficient variation in emission line strengths and ratios, or even better, first estimate the redshift empirically and fit the colours with templates at the best-fit redshift to calculate the K-correction and various physical parameters.
\end{abstract}


\section{Introduction}

Due to the well-known limitations of spectroscopic observations, ongoing and upcoming wide area cosmological surveys, such as DES (Dark Energy Survey, \citealt{DES}), HSC (Hyper Suprime-Cam Subaru, \citealt{Subaru}), KiDS (Kilo-Degree Survey, \citealt{KiDS}) LSST (Large Synoptic Survey Telescope, \citealt{LSST}) and Euclid (\citealt{Euclid}), depend highly on reliable photometric redshifts to recover the cosmological parameters with sufficient accuracy. Photometric redshift evaluation methods can be split in two groups: the template-based methods that rely on representative libraries of model spectra to fit the spectral energy distributions of galaxies (see \citealt{Benitez2000}; \citealt{Budavari2001}; \citealt{Csabai2003}; \citealt{Ilbert2006}; \citealt{Brammer2008} and references therein) and the empirical methods that apply machine learning techniques to infer the hidden correlation between photometric data and redshift. There are numerous new machine learning based photo-z estimators in development, with a large variety of approaches for providing the most precise measurements of cosmological parameters \citep{Sadeh2016, Speagle2016, Cavuoti2017, Graham2018}. Current, state-of-the-art photo-z codes use empirical methods, often combined with a Bayesian approach \citep{Benitez2000, Dahlen2010, Tanaka2015}, to derive the probability distribution $p(z_{\textnormal{phot}})$ for each galaxy, and the $N(z)$ distribution function for the total ensemble. Combined with regularization techniques \citep{Bordoloi2010} and folding in galaxy clustering information (\citep{Morrison2017, Scottez2018}), photo-z uncertainties can be well controlled and propagated into the final cosmological parameter uncertainties. For a more complete overview on photometric redshift evaluation methods see \cite{Salvato2019}.

One major difficulty of the photometric redshift evaluation methods is that the current and future spectroscopic surveys are incomplete in magnitude, redshift and physical properties \citep{Cooper2006}, along with the fact that spectroscopic surveys to date do not sample the full colour space of galaxies in the deeper, Euclid-like surveys, and thus the colour-redshift relation is not fully constrained with existing spectroscopy, which makes the photometric redshift calibration challenging \citep{Masters2017}. This issue was investigated in detail by \cite{Beck2017} who constructed two separate sets of galaxy catalogues to model the biases present in the photo-z estimations either due to the lack of training set coverage in the feature space compared to the test set or due to the mismatch between photometric error distributions of the two samples. They found that the mismatch between distributions could be adequately handled by template fitting and empirical methods as long as training sample coverage is sufficient, while the issue of spectroscopic coverage mainly lowers the precision of the local machine-learning based estimations, but the calibration of template-based techniques is also affected. They also note that in case of single instrument spectroscopic observations the results cannot be tested on poorer quality photometry from another survey, which in turn means that more than the feature-space coverage has to be taken into account to address the photo-z issues. Moreover, projections for cosmic shear measurements considering Euclid statistics indicate two requirements on photo-z for precision weak lensing: the precision for each object should be better than $0.05(1+z)$ and the redshift bias, or the true mean redshift of objects in each photo-z bin must be known with at least $\Delta \overline{z} = 0.002(1+z)$ precision (\citep{ZhanKnox2006, Bordoloi2010}. To estimate and understand the physical reasons behind photo-z uncertainties and bias beyond the limited training sets, photometric noise and other observational effects, we should investigate the distribution of model galaxies in the colour-colour space with as much detail as possible. This will not only us help get an insight into what can go wrong with empirical techniques but also help improve spectroscopic targeting strategies to compile spectroscopic training sets with optimized colour space coverage.

\subsection{Template-based and empirical photometric redshift estimation}

Template-based photometric estimation works by fitting model spectra to observed broad-band colours. Synthetic magnitudes from the models are computed by convolving the templates with transmission curves of the filters used for the observations. While this approach is physical in the sense that spectral templates can be derived from theoretical models with known physical input parameters, template-based photo-z usually performs significantly worse than empirical methods due to differences between the models and real galaxies and discrepancies between synthetic and real magnitudes \cite{Salvato2019}.

Empirical photometric redshift estimation methods are based on the application of a training set (galaxies with both known spectroscopic redshift and broad-band colours) and machine learning methods that interpolate from the properties of the training set galaxies to galaxies with available photometry but no spectroscopic redshift. It has become clear, however, that empirical methods can be severely limited by the availability of a training set with reliable spectroscopic redshift measurements. The main limiting factors of training sets are the sparse coverage of the multi-dimensional colour-colour space that photometric galaxies span and the unavoidable necessity to extrapolate to faint magnitudes where spectroscopic observations are not feasible. The training sets usually cover only the brighter end of observations, thus extrapolation to faint objects -- which might have got different spectral features than brighter ones -- is also often necessary. The quality of the latter is hard to measure. Template-based methods could overcome these limitations, but they lag behind empirical techniques due to the problems mentioned above.

\subsection{Related work}

The large amount of flux-calibrated optical spectra accumulated by the SDSS opened up new ways to test the various spectral synthesis models and theories. The connection of stellar population synthesis and radiative-transfer models allowed the modelling of emission lines on top of stellar continua, most notably \textsc{PÉGASE} \citep{Fioc1997} and \textsc{BPASS} \citep{Eldridge2012}, but the photoionization models of these software packages introduces a large set of free parameters that describe the distribution and composition of the ISM. The number of these parameters can be reduced by theoretical or empirical assumptions, i.e. assuming a certain symmetry for the ISM \citep{Stasinska1984} or common ionization spectrum for all gas clouds \citep{Ferland2017}, or by estimating the priori distribution of model parameters using large ensembles of models with observations, as for instance, in case of the \texttt{LE PHARE} code \citep{Ilbert2006}.

Beyond existing theoretical methods, the large amount of observed galaxy spectra allow for an empirical route to generate realistic emission lines on top of stellar continua derived from population synthesis models. \cite{Yip2004} demonstrated that the spectra of SDSS galaxies form a sequence which, in leading order, can be parametrized by a single parameter by expressing the spectra on a basis derived by principal component analysis. \cite{Gyory2011} and \cite{Beck2016a} showed that correlations exist between the principal components (PCs) of stellar continua and emission line $\log EW$s. \cite{Beck2016a} also gave a recipe to generate realistic emission lines on model stellar continua by leveraging these correlations. The correlations are not very tight: instead of finding scaling relations or fitting formulae between line and continuum PCs, one has to use probability distributions to describe emission line strengths and correlate the parameters of the distributions with PCs of the continuum. In our paper, we extend their approach for a wider set of galaxies and create a realistic mock catalogue on a purely empirical basis, that matches the observed distributions.

\subsection{Analysing the effect of emission lines}

The principal goal of this paper is to assess the effect of strong emission line galaxies, and the lack of sufficient training set coverage of them, on photometric redshift estimation. For this purpose we generate model galaxy spectra with infinite signal-to-noise ratio that match the luminosity, redshift, colour and emission line distributions of a spectroscopic sample. 

The advantage of the outlined technique is clearly its simplicity over approaches which require detailed modelling of the gas and sources of excitation in galaxies to determine line strengths. Consequently, instead of trying to find the right priors on population synthesis and photoionization models to generate realistic simulated catalogues, we start from observations and build a stochastic model that can generate physically meaningful model spectra by linear combination of stellar continuum eigenspectra and emission line eigenvectors derived from logarithmic equivalent widths. The method starts by fitting the stellar continuum and emission lines of the observed spectra, then running a principal component analysis on both. After classifying the fitted models according to a few parameters, such as single-band luminosity and the first two continuum PCs, we describe the distribution of both continuum and line PCs in each class with Gaussian mixture models (GMM). By drawing samples from a prescribed luminosity function and PCs from the Gaussian Mixture Models, we can generate very realistic spectroscopic and photometric catalogues with arbitrary emission line contribution and photometric noise. This opens the way to test empirical photo-z methods on samples with and without these effects to compare their outcome. We intentionally limit ourselves to the simplest photo-z techniques so that the complexities of the more advanced codes do not obscure the lessons we can learn from turning emission lines and photometric error on and off.

\noindent The paper is structured as follows. In Sec.~\ref{sec:data}, we introduce the initial observed data set we start from and outline the continuum and emission line fitting of the spectroscopic sample. We explain the PCA-based stochastic method for generating realistic mock spectra in Sec.~\ref{sec:mock}, then investigate the basic properties and distributions along with the influence of emission lines on these in Sec.~\ref{sec:mock_eval}. Following a brief summary of the empirical photo-z methods we test in Sec.~\ref{sec:methods}, we continue by presenting the results of the analysis in Sec.~\ref{sec:results} and we conclude the paper in Sec.~\ref{sec:conclusions}. The python code that was used for the analysis is available on the GitHub page of the author: \url{https://github.com/Csogeza/EmiPhotoZ}

\section{Data Set and Data Processing}
\label{sec:data}

In order to match the spectroscopic mock catalogue to observations, we use a data set derived from the SDSS main galaxy spectroscopic sample. Although shallow in redshift, the advantage of using SDSS over newer, deeper data sets is its relatively unbiased target selection method and excellent colour-space coverage. The obvious consequence of using SDSS is restricting ourselves to low redshifts of $z < 0.35$ and shallow photometric depths of $m_r < 17.77$. On the other hand, we only use the SDSS sample as a starting point to generate the mock catalogue and our method allows for reasonable extrapolation to depths where spectral evolution effects are still considered small.

\begin{figure}
\includegraphics{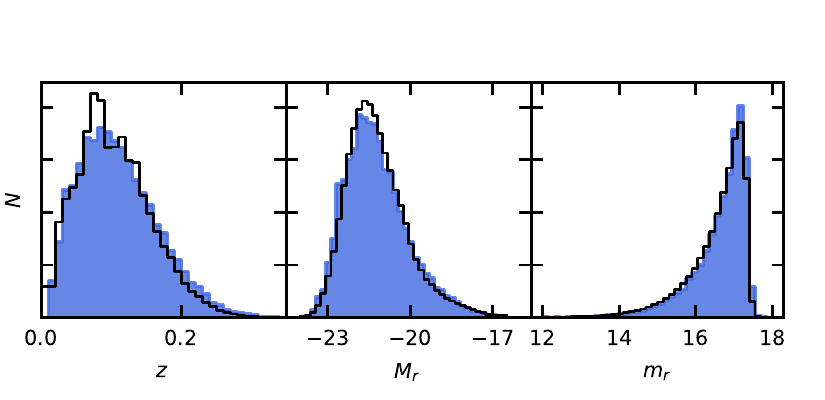}
\caption{\textbf{Left:} Normalized redshift distributions of the original SDSS DR7 spectroscopic main galaxy sample (solid line) and the mock catalogue (blue histogram) generated by our method. \textbf{Middle:} Normalized distribution of absolute $r$-band Petrosian magnitudes of the original SDSS sample (solid line) and the mock catalogue (blue histogram). \textbf{Right:} Normalized apparent $r$-band Petrosian magnitude of the original SDSS sample (solid line) and the mock catalogue (blue histogram). The minor differences are thought to be primarily to the large scale structure that our model does not take into account.}
\label{fig:catalog_hists}
\end{figure}

The SDSS main spectroscopic galaxy sample consists of a mixture of blue and red galaxies at low redshifts and mostly luminous red galaxies (LRGs) above $z > 0.1$. The left panel of Fig.~\ref{fig:catalog_hists} shows the normalized redshift distribution, as well as the distributions of absolute and apparent $r$-band magnitudes of the initial empirical sample we attempt to match with the mock catalogue.

We divided the original SDSS sample into ``strong emission line'' and ``other'' galaxies, the latter to include all galaxies that have got Hydrogen emission only or show no prominent emission lines. Galaxies showing no emission lines will be termed ``passive'' and the rest ``weak emission line'' galaxy. The sub-sample of emission line galaxies was selected based on the criterion that all nebular emission lines of Tab.~\ref{tab:linelist} should be detected at a signal-to-noise ratio larger than $3$. Passive and weak line galaxies were treated as a continuum based on the strength of the H$\alpha$ line.

\begin{table}
\label{table_lines}
\begin{center}
\begin{tabular}{l l | l l }
	\hline
	Line & $\lambda_\textnormal{vac}$ [\AA] & Line & $\lambda_\textnormal{vac}$ [\AA] \\ \hline \hline
	H$\alpha$		& 6565	& H$\beta$		&	4863 \\ 
	\ion{S}{ii}		& 6718	& \ion{S}{ii}	& 6733 \\
	\ion{O}{ii}		& 3727	& H$\gamma$		& 4342 \\
	\ion{O}{iii}	& 5008	& \ion{O}{iii}	& 4960 \\
	\ion{N}{i}		& 6586	& \ion{N}{ii}	& 6556 \\
	\hline
\end{tabular}
\caption{The list of emission lines used to define the ``emission line'' galaxy sample. All of these lines must be detected at $3\sigma$ level for a galaxy to be classified as ``emission line'' galaxy.}
\label{tab:linelist}
\end{center}
\end{table}

The data processing steps for ``strong emission line'' galaxies are outlined in the top panel of Fig.~\ref{fig:flowchart}. Each step is explained in details in the following sections.

\subsection{Continuum PCA}
\label{sec:cont_pca}

To be able to match the observed data with infinite signal-to-noise ratio models, we first correct the flux calibrated spectra for foreground extinction based on the law of \cite{Calzetti2000} and \cite{ODonnell1994} using the dust map of \cite{Schlegel1998}, then fit the intrinsic extinction, velocity dispersion and stellar continuum of the empirical data set with non-negative linear combinations of stellar population templates generated with the code of \cite{Bruzual2003} and  \cite{Tremonti2004}, as described in \cite{Beck2016a}. Intrinsic extinction is modelled using the prescriptions of \cite{Charlot2000}, and extinction is re-applied to the model continua after fitting. Before further analysis the obtained model spectra are then resampled to a common wavelength grid, then normalized according to the procedure detailed in \cite{Beck2016a} by setting the average of median fluxes in various featureless wavelength ranges to 1, to account for the flux differences among the spectra.

\begin{figure*}
\includegraphics[width=\columnwidth]{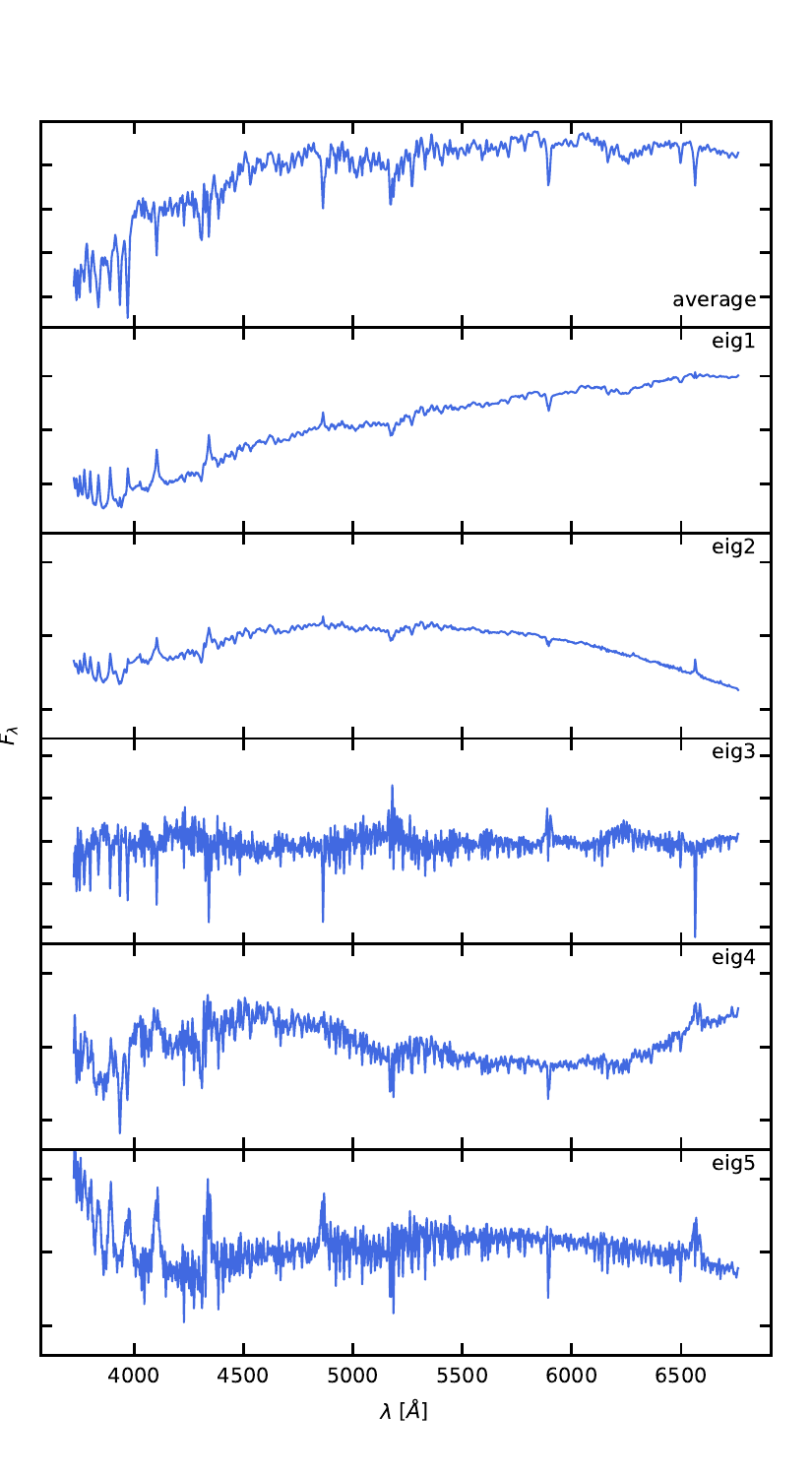}
\includegraphics[width=\columnwidth]{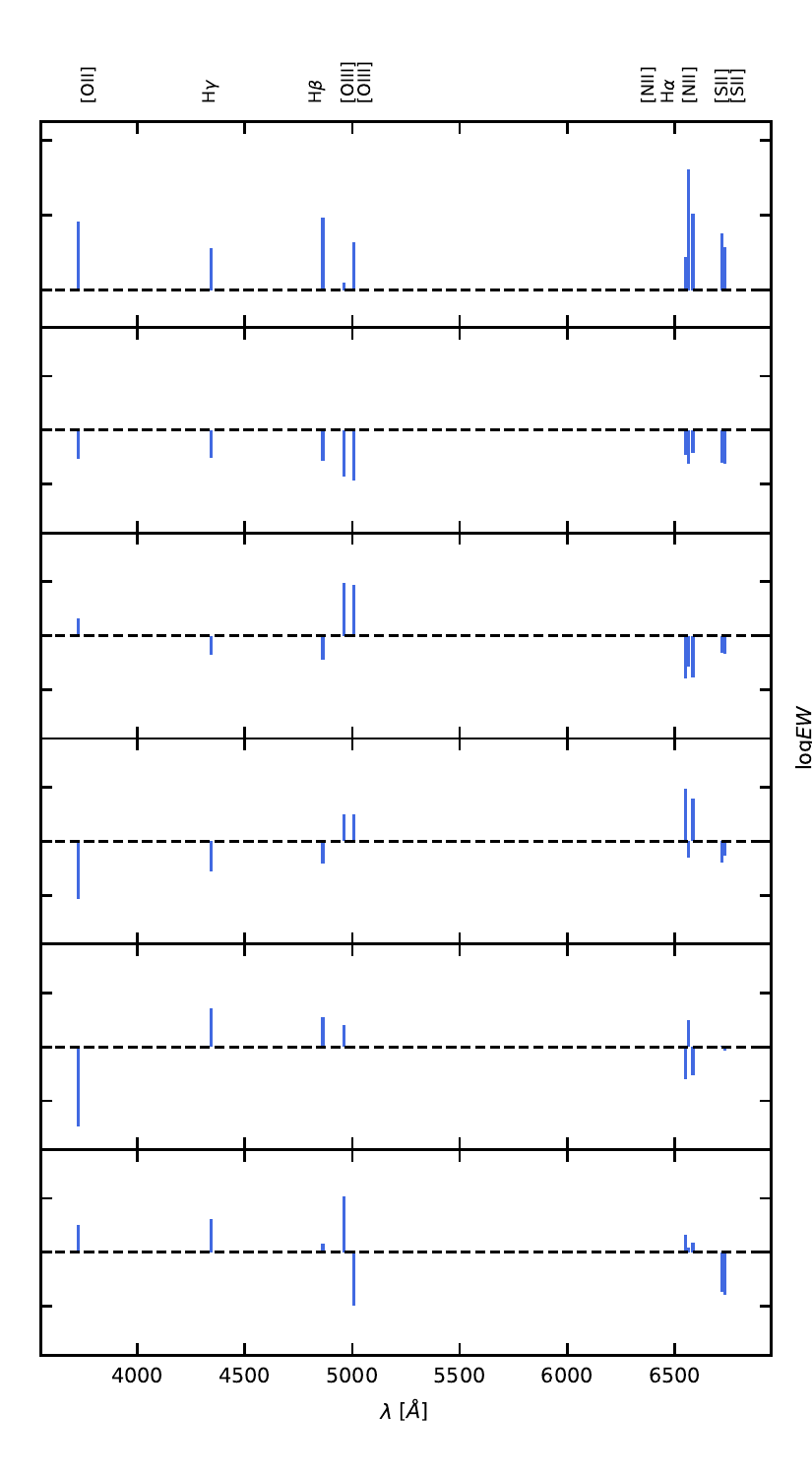}
\caption{Stellar continuum (left) and line $\log EW$ (right) average vectors (first row) and eigenvectors derived from models fitted to the spectra of the SDSS main galaxy sample. While the continuum eigenspectra account for all galaxies, emission line eigenvectors are only computed from the emission line galaxy sample.}
\label{fig:eigs}
\end{figure*}

To reduce the dimensionality of the fitted continua, we run Principal Component Analysis on the model flux vectors and plot the resulting average spectrum and the first five eigenspectra in the left panels of Fig.~\ref{fig:eigs}. Since the third and higher eigenspectra appear to be sensitive to absorption features only, we expect that broadband colours of the models are mostly determined by the first two eigenspectra. We note, that even though we use the linear combination of only 10 Bruzual--Charlot templates to fit the stellar continua of the observed spectra, due to the non-linearity of intrinsic extinction, we can do the PCA without restricting ourselves into a linear subspace.

\begin{figure}
\includegraphics{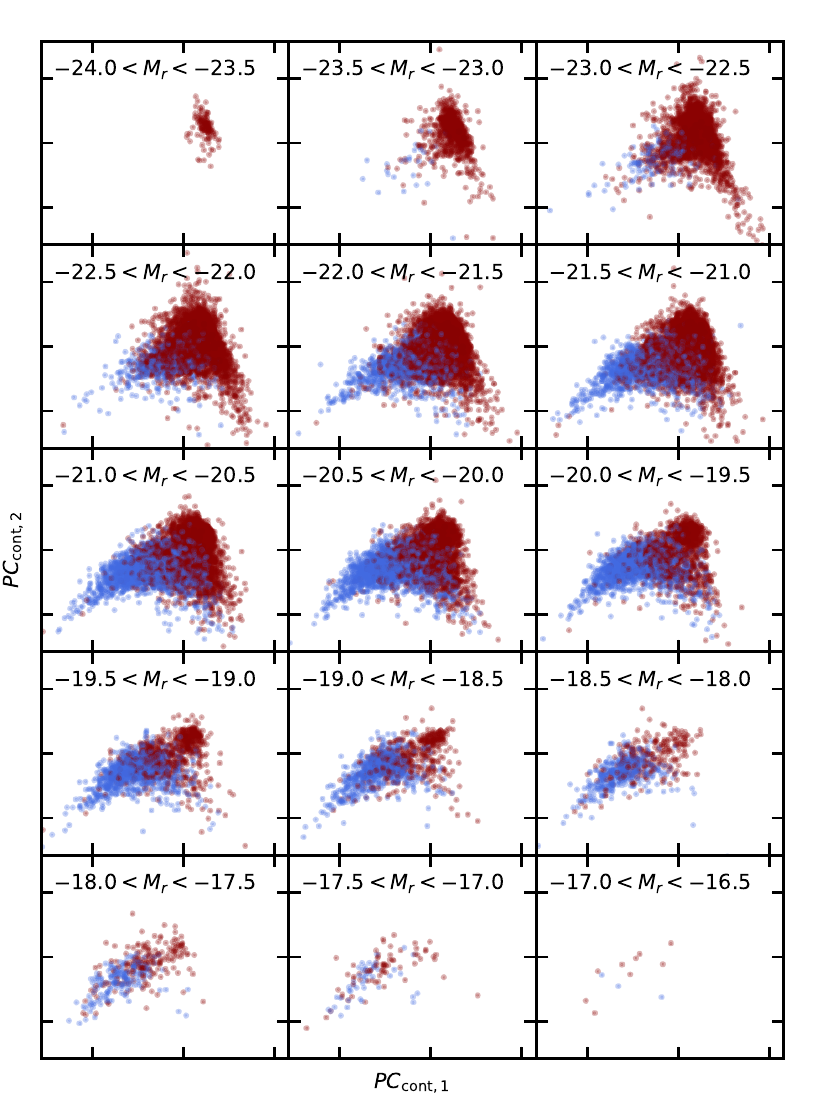}
\caption{The distribution of the first two continuum principal components in different absolute magnitude bins between $-16.5 < M_r < -23.5$. Blue dots represent galaxies with emission lines detectable at $3\sigma$ level, whereas red dots are ``passive'' galaxies, i.e. galaxies with no emission lines, weak emission lines or Hydrogen lines only.}
\label{fig:cont_pcs}
\end{figure}

For the purpose of mock catalogue generation, we split the sample into absolute magnitude intervals, as listed in Tab.~\ref{tab:abs_mag_intervals}, and plot the distribution of the first two principal components in Fig.~\ref{fig:cont_pcs} for each absolute magnitude bin, with different colours for ``strong emission line'' and ``passive + weak emission line'' galaxies. In each of the absolute magnitude intervals, we model the distribution of the first five continuum principal components with multivariate Gaussian mixture models. The number of components used for the fitting of the distributions are determined based on how well the samplings from the resulting GMMs reproduce the original point clouds in each luminosity bin. To measure this, we fitted the subsets with GMMs with the number of components varying between 2 and 25, then compared the resulting sample distributions of these models for the different principal components using the two-sample Kolmogorov–Smirnov statistic \cite{Massey1951}. For the luminosity bins with the lowest galaxy count the distributions are best modelled with two mixture components, with this number increasing for bins with more numerous data entries. We found that the KS statistic of the GMM fit did not improve above $N=11-15$ for the subsets with the highest amount of datapoints, thus we limited the number of mixture components used for the fitting to 15, which was only used for the most populous luminosity bins, to avoid the possible issue of overfitting.


\begin{table}
\begin{center}
\begin{tabular}{c | c c }
	\hline
	Absolute magnitude & Strong e. l. & Passive + Weak e. l. \\
	\hline
	\hline
	-24.0 <M< -23.5 & 0 & 128 \\ 
	-23.5 <M< -23.0 & 30 & 1369 \\ 
	-23.0 <M< -22.5 & 187 & 5591 \\ 
	-22.5 <M< -22.0 & 738 & 8034 \\
	-22.0 <M< -21.5 & 1654 & 8918 \\
	-21.5 <M< -21.0 & 2313 & 7901 \\
	-21.0 <M< -20.5 & 2642 & 5163 \\
	-20.5 <M< -20.0 & 2324 & 3066 \\ 
	-20.0 <M< -19.5 & 1685 & 1522 \\ 
	-19.5 <M< -19.0 & 1051 & 707 \\
	-19.0 <M< -18.5 & 647 & 432 \\
	-18.5 <M< -18.0 & 375 & 246 \\
	-18.0 <M< -17.5 & 168 & 183 \\
	-17.5 <M< -17.0 & 46 & 70 \\
	-17.0 <M< -16.5 & 3 & 9 \\
	\hline
\end{tabular}
\end{center}
\caption{The number of strong emission line and passive/weak emission line galaxies in the various absolute magnitude intervals. As it is expected, the brighter end of the galaxy sample consists of mostly passive or weak emission line galaxies, owing to the fact that these galaxies are mostly large mass ellipticals, whereas the distribution among the two classes is more even for the fainter galaxies.}
\label{tab:abs_mag_intervals}
\end{table}

\subsection{Line PCA}
\label{sec:line_pca}

\begin{figure}
\includegraphics[width=\columnwidth]{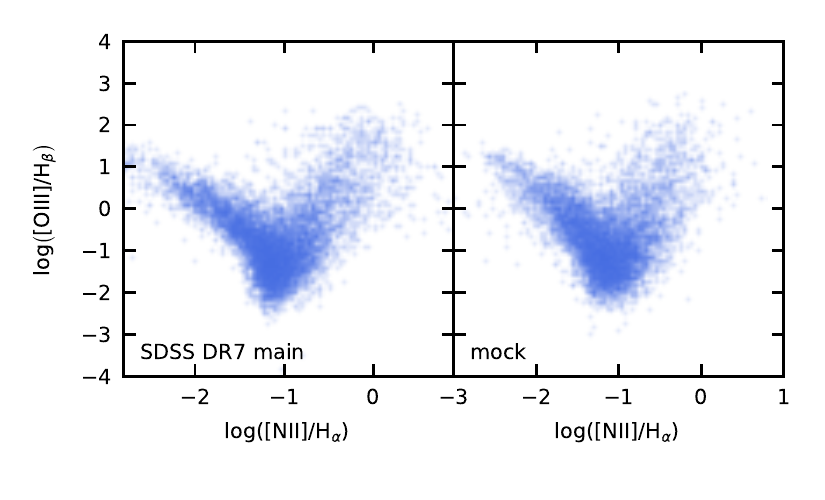}
\caption{The Baldwin--Philips--Terlevich (BPT) diagram of the original SDSS DR7 main galaxy sample and the mock catalogue generated by our technique. Only galaxies with emission line SNR better than 3 are plotted in the left panel. Although higher randomness is visible in the mock sample, it matches observations reasonably well.}
\label{fig:bpt}
\end{figure}

To measure the equivalent width of emission lines, we applied the noise-limited technique developed for fitting strong and asymmetric AGN lines by \cite{Beck2016a}. For those galaxies which have got emission lines measurable at least at $3\sigma$, we plot the \cite{BPT} diagnostic diagram in the left panel of Fig.~\ref{fig:bpt}. The BPT diagram shows the strengths of the strongest nebular lines normalized to the closest Hydrogen lines. We refer the reader to \cite{Kewley2013} to learn more about the BPT diagram of SDSS galaxies. Any mock catalogue generation technique that accounts for emission lines should, at the minimum, reproduce the BPT diagram to match line ratios.

\begin{figure}
\includegraphics{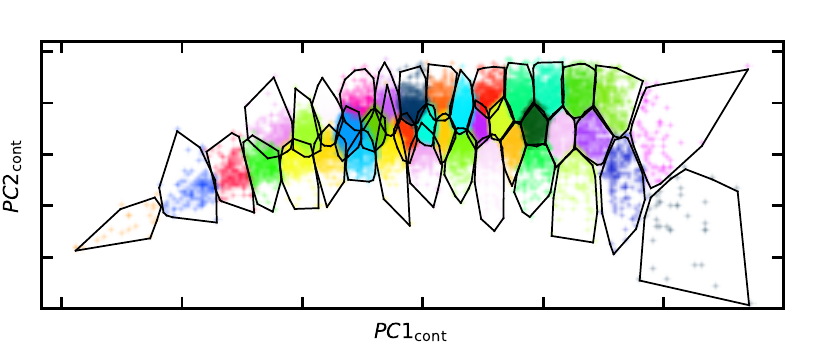}
\includegraphics{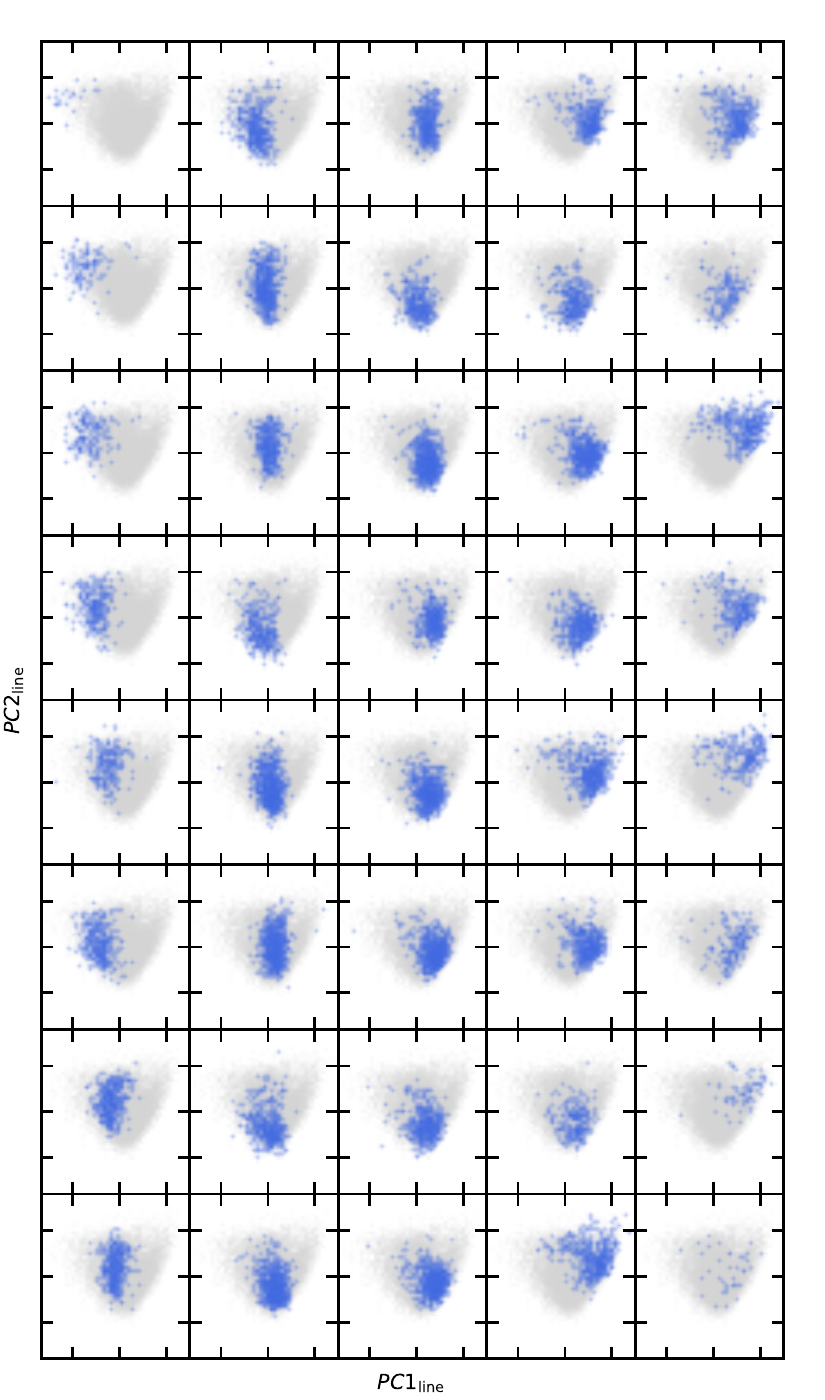}
\caption{\textbf{Top:} Clusters derived from continuum principal components by running the $k$-means algorithm over the first five PCs. The black outlines show the projection of the convex hull of the clusters to the first two PCs. The overlap between the clusters in the first two principal components is minimal due to how PCA works. \textbf{Bottom:} Distribution of line principal components in each continuum cluster plotted over the distribution of line PCs of the entire emission line galaxy sample. Continuum clusters are ordered by the first component of their centroids, from left to right, top to bottom. The distribution of line PCs clearly depends on where the corresponding continuum is located in the PC space, but no tight correlations are observable.}
\label{fig:cont_clusters}
\end{figure}

We form data vectors from the logarithms of equivalent widths of each line listed in Tab.~\ref{tab:linelist}. Taking the logarithm of equivalent width ensures that line ratios will persist after dimensionality reduction by PCA. We compute the principal components for the entire ``emission line'' sub-sample and plot the average $\log EW$ vector and the first five eigenvectors in the right panel of Fig.~\ref{fig:eigs}. The first emission line eigenvector sets the primary line ratios and it captures the majority of differences between the emission of different types of galaxies. The second eigenvector captures mainly the correlation among Oxygen lines, thus it is expected to correlate with the vertical position of the galaxies on the BPT diagram. The third eigenvector mainly contains information about the [NII] and [OIII] emission strengths and the fourth vector sets the strength of the Balmer series lines, which primarily determine the horizontal location on the BPT diagram. The last shown eigenvector most probably contains noise terms, indicated by the large difference between the strengths of [OIII] lines.

\subsection{Continuum--line correlations}
\label{sec:contlinecorr}

To investigate how the number fraction of galaxies from the various emission types depends on the absolute brightness, we plotted the distributions of the first two derived continuum PCA coefficients of the compiled galaxy set in the various brightness ranges in Fig.~\ref{fig:cont_pcs}. It is clear from Fig.~\ref{fig:cont_pcs} that the bright-end of the sample consist of passive Luminous Red Galaxies (LRGs) whereas emission line galaxies, both star-forming ones and galaxies with active nuclei tend to have somewhat lower luminosity. To measure the probability of a galaxy having strong emission lines, we simply bin galaxies by the first two continuum principal components in each absolute magnitude bin and take the ratio of ``strong emission line'' and ``passive + weak emission line'' galaxies. In Sec.~\ref{sec:stronglines}~and~\ref{sec:weaklines}, we are going to use these luminosity dependent distributions of the continuum principal components and the probabilities of having all lines listed in Tab.~\ref{tab:linelist} to generate the emission lines on top of stellar continua of the mock spectrum catalogue.

We further organize ``strong emission line'' galaxies, regardless of their luminosity, into automatically determined classes based on the first five continuum principal components. The results from $k$-means clustering with the Euclidean metric and $k = 40$ are shown in Fig.~\ref{fig:cont_clusters}. When choosing the number of clusters for the modelling we aimed to cut the complete dataset into smaller subsets which can represent distinct subclasses of galaxies, while keeping the number of clusters low enough that each cluster will contain at least 20-30 galaxies. The resulting clusters are sufficiently small for the stellar continua in each of them to be very similar to each other in both broad-band colours and absorption features, thus the average spectrum in each cluster represents the cluster well. We are going to model the correlations between the stellar continua and emission lines on a per-cluster basis. Here we make the assumption that emission lines show stronger correlations with the shape of the stellar continuum than with the intrinsic luminosity of the galaxy, hence we can avoid further binning by absolute magnitude. Consequently, the absolute magnitudes of mock galaxies determine the probability of presence of strong emission lines but not equivalent widths and line ratios which are solely determined by their correlations to the stellar continuum.

To quantify the correlations between the stellar continua and strong emission lines, we consider the distribution of emission line principal components within each continuum k-means cluster. In Fig.~\ref{fig:cont_clusters}, we plot the distribution of the first two line principal components for each continuum clusters. Similarly to modelling the distribution of continuum PCs with Gaussian mixture models, we also use GMMs to describe the distribution of line principal components within each of these continuum clusters.

\begin{figure}
\includegraphics{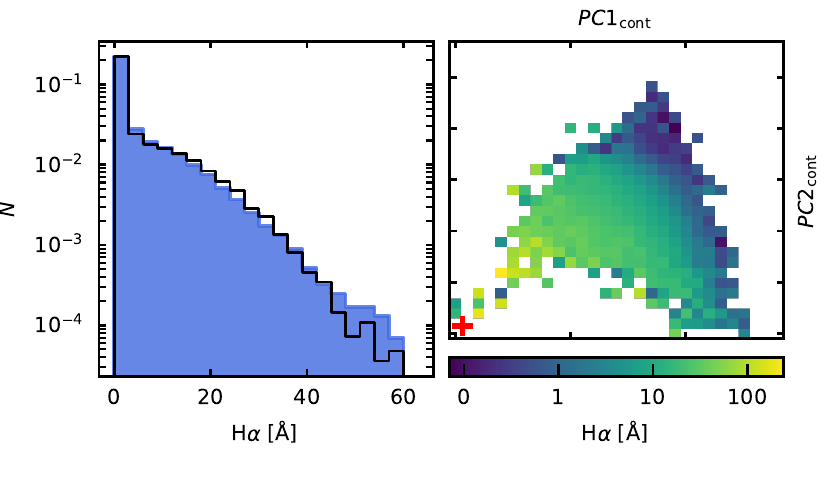}
\caption{\textbf{Left:} Distribution of H$\alpha$ line equivalent widths for passive galaxies and galaxies with detectable H$\alpha$ but no significant nebular emission lines, in case of SDSS (black curve) and the mock catalogue (blue histogram). The large peak at zero is due to passive galaxies with no detectable H$\alpha$ line. \textbf{Right:} Average equivalent width of the H$\alpha$ line (colour coded) as a function of the first two continuum principal components. The red cross marks the locus from which distance $d$ is measured, c.f. Eq~\ref{eq:halpha} and Fig.~\ref{fig:halpha_fit}.}
\label{fig:halpha_lines}
\end{figure}

In case of ``passive + weak emission line'' galaxies, we focus on correlations between the continuum principal components and the strength of the H$\alpha$ line only. Fig.~\ref{fig:halpha_lines} shows the equivalent width of H$\alpha$ as a function of the first two continuum principal components regardless of absolute magnitude. The independence of H$\alpha$ equivalent widths from absolute magnitude results from the fact that most of the weak emission and passive galaxies are quiescent star forming galaxies, for which a relation between the strength of emission and continuum shape is expected from the correlation between the Balmer-line strengths and the 4000\AA\hspace*{1pt} break \citep{Kauffmann2003}. The plotted region corresponds to the distribution of red dots in Fig.~\ref{fig:cont_pcs}. One can find that the equivalent width of H$\alpha$ emission shows a linear dependence on the first two continuum principal components in the form of
\begin{equation}
	EW_{\textrm{H}\alpha} = \alpha \sqrt{(PC_1 - a)^2  + (PC_2 - b)^2} + \beta,
	\label{eq:halpha}
\end{equation}
where $a$ and $b$ are reference principal component coefficients, which describe a reference point, from which the Euclidean distance is measured in the PC space (in Fig.~\ref{fig:halpha_lines}). This reference spectra is denoted by the red cross in the figure. The $\alpha$ and $\beta$ parameters describe the linear dependence of equivalent width on the defined distance measure.

\subsection{Photometric errors}
\label{sec:photo_error}

\begin{figure}
\includegraphics{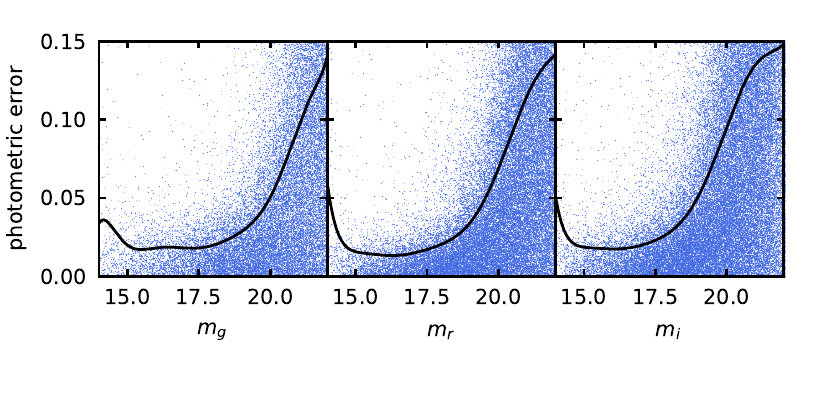}
\caption{Estimation of photometric error from repeated observations. The dots show the absolute difference of apparent Petrosian magnitudes as a function of magnitude, whereas the solid curve is a fit to the distribution in magnitude bins of $0.008$. The upturn at bright magnitudes is due to low galaxy counts and repeatability problems of image reduction.}
\label{fig:phot_error}
\end{figure}

To be able to add realistic noise to mock apparent magnitude, instead of using the error estimates of the SDSS photometric reduction pipeline, we consider repeated observations. Fig.~\ref{fig:phot_error} shows the distribution of the differences between $r$-band magnitudes $m_r$ of two observations of the same galaxy. We consider the error to be Gaussian in bins of $\Delta m = 0.025$ and fit the $1\sigma$ error as a function of $m$ with a 10$^\textrm{th}$ order polynomial in each band. We then calculated the final photometric error for each band as the sum of the polynomial fit and a random term, which was drawn from a normal distribution with a scatter of 1$\sigma$ of the fits. With the choice of a very high degree polynomial we aimed to capture the leading trend in the photometric error without assuming any strict dependency on the apparent brightnesses of galaxies. Equivalent trends can be captured simply by calculating the average curve of the data or by applying generalized additive model regression to the dataset, thus the exact choice of the models does not alter the results obtained through the fitting. Thus, for simplicity, we have chosen to model the photometric errors with a high degree polynomial trend. Since we only aim to analyse the randomizing effect cause by the photometric errors qualitatively, a simple modelling reproducing the general trend in the errors is sufficient for our purposes. However, we note, that in the case of using this catalogue for the purposes of investigating a real sample of galaxies, the photometric errors should inevitably modelled in more detail, to match the photometric error distribution of the real sample correctly, either though more complex models or by taking into account the possible correlations with other observation parameters as well instead of a single apparent magnitude. 

\section{Mock catalogue generation}
\label{sec:mock}

\begin{figure}
\includegraphics{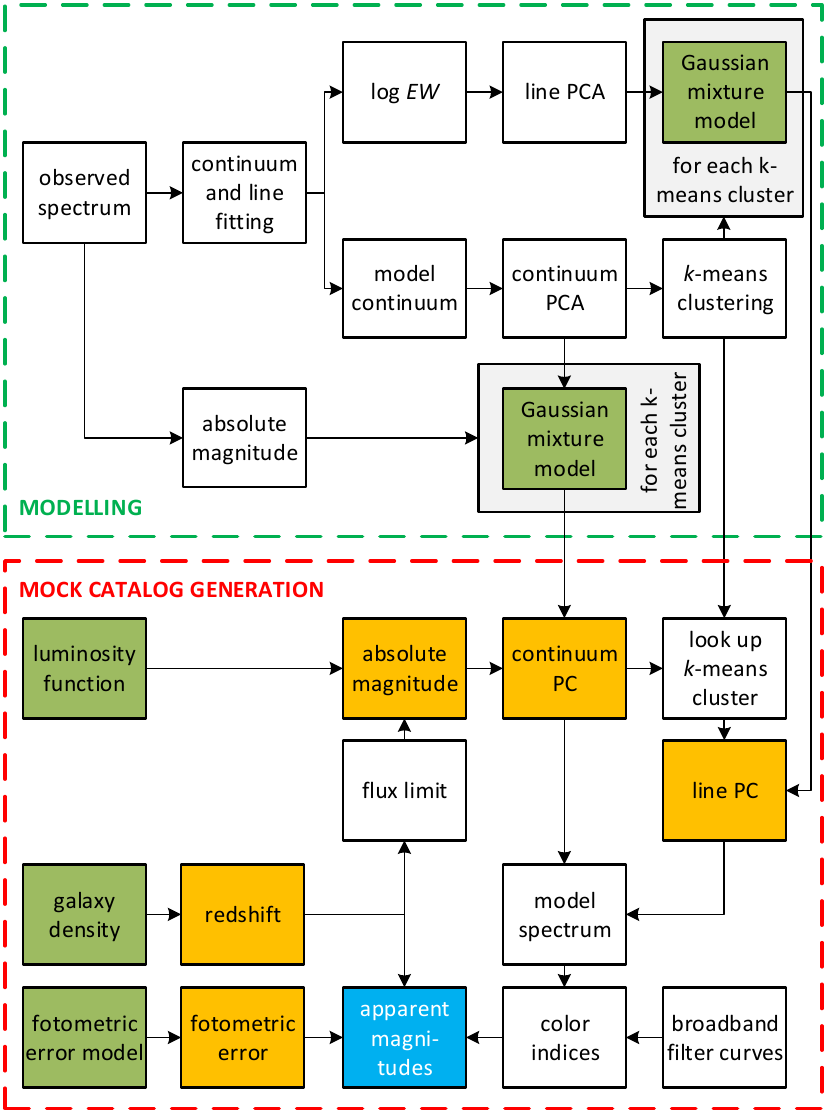}
\caption{The method of mock catalogue generation. The upper panel shows how we model the continuum and emission lines of the observed galaxies by Gaussian mixture models of the continuum and line principal components. The lower panel explains how model spectra are generated by drawing samples from the theoretical distributions (for absolute magnitude and redshift) and fitted stochastic models (for continuum and line principal components and the photometric error). This flowchart applies to strong emission line galaxies only, see text for details about passive and weak emission line galaxies.}
\label{fig:flowchart}
\end{figure}

We generate our mock catalogue to match observational properties of the SDSS~DR7 main spectroscopic sample. Continuum and emission line generation, rather than starting from physical properties, are based on randomized principal components as introduced by \cite{Beck2016a}. The main steps of the algorithm are illustrated in the bottom half of Fig.~\ref{fig:flowchart} for ``strong emission line'' galaxies. The algorithm for ''weak emission line + passive'' galaxies is somewhat simpler as only the H$\alpha$ line and its correlation of the stellar continuum is taken into account.

\subsection{Stellar continuum generation}
\label{sec:gencont}

Mock catalogue generation starts by drawing SDSS $r$-band absolute magnitudes and redshifts from theoretical distributions. We sample SDSS $r$-band luminosities from the standard function of \cite{Schechter} with parameters $\alpha = -1.23$ and $M_* = -21.53$. To generate the redshifts, we assume constant galaxy density in a standard $\Lambda$CDM setting with parameters $H_0 = 69.32$~km~s$^{-1}$~Mpc$^{-1}$, $\Omega_M = 0.287$ and $\Omega_\Lambda = 0.713$. Only galaxies passing the flux limit are kept\footnote{To simplify catalogue generation, when applying the flux limit, we do not take emission lines and photometric error into account.}. The panels of Fig.~\ref{fig:catalog_hists} show the distribution of randomly generated absolute magnitudes and redshifts and the corresponding apparent magnitudes for the mock catalogue. Our method can reproduce the redshift distribution well, although the effect of the large-scale structure is not accounted for.

To construct realistic stellar continua based on the eigenspectra derived from PCA, first we consider the absolute magnitude of each mock galaxy and look up the absolute magnitude bin it belongs to, c.f. Tab.~\ref{tab:abs_mag_intervals} and Sec.~\ref{sec:cont_pca}. Then, based on the Gaussian mixture model corresponding to the absolute magnitude bin, we randomly draw coefficients and calculate the linear combination of the continuum eigenspectra. 

\subsection{Emission line generation}

When adding emission lines to the mock stellar continua, we distinguish ``strong emission line'' and ``weak emission line + passive'' galaxies. The right mixture of the groups and the correlations between the lines and the stellar continuum are essential to generate a realistic mock catalogue for the purpose of photo-z testing. 

Depending on the continuum coefficients drawn in the previous step (see Sec.~\ref{sec:gencont}) we randomly classify mock galaxies as ``strong emission line'' or ``weak emission line + passive''. The probability of a galaxy having strong emission lines is taken as described in Sec.~\ref{sec:contlinecorr} and visualized in Fig.~\ref{fig:cont_pcs}, where we plotted the first two principal components of the SDSS galaxies in absolute magnitude bins, colour-coded by the presence of strong emission lines. If the mock galaxy is classified as ``strong emission line'', we generate lines with the algorithm described in Sec.~\ref{sec:stronglines}, whereas in case of the rest of the galaxies, we only add the H$\alpha$ line as discussed in Sec.~\ref{sec:weaklines}.

\subsubsection{Strong emission line galaxies}
\label{sec:stronglines}

\begin{figure}
\includegraphics{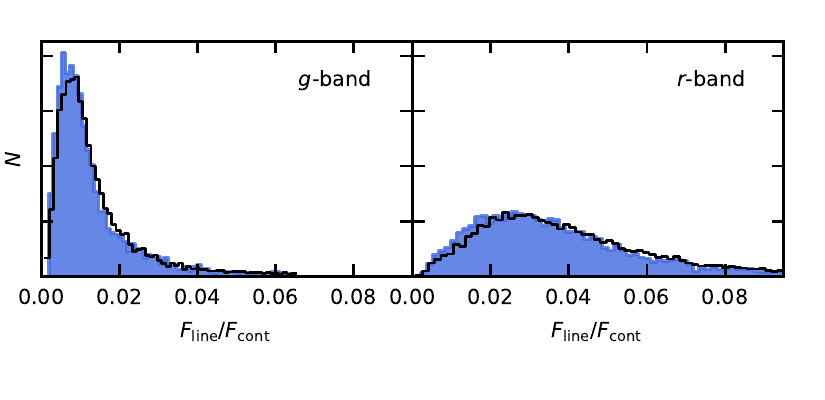}
\caption{The distribution of line-to-continuum flux ratios in the SDSS $g$ and $r$ photometric bands for the SDSS DR7 emission line galaxy sample (solid line) and the mock catalogue (blue histogram) generated by our method. Contributions of lines to the total flux in these bands can reach as high as a few per cent.}
\label{fig:line_flux_ratio}
\end{figure}

If a mock galaxy is selected as ``strong emission line'', based on the random continuum coefficients, it is classified into one of the 40 continuum k-means clusters, see Fig.~\ref{fig:cont_clusters}. In Sec.~\ref{sec:line_pca}, we detailed how the distribution of principal components of emission line $\log EW$s are described using Gaussian mixture models on the set of principal component coefficients. Based on the continuum k-means cluster the mock galaxy falls into, we draw its emission line coefficients randomly from the corresponding Gaussian mixture model. Lines are then added on top of the mock galaxy continua as random linear combinations of the line eigenvectors derived in Sec.~\ref{sec:line_pca}. To examine how well the empirical method reproduces the flux excess caused by emission lines, we plotted the distribution of line flux-- continuum flux ratio for the mock catalogue and the galaxy sample in Fig.~\ref{fig:line_flux_ratio}.

\subsubsection{Weak line and passive galaxies}
\label{sec:weaklines}

\begin{figure}
\includegraphics{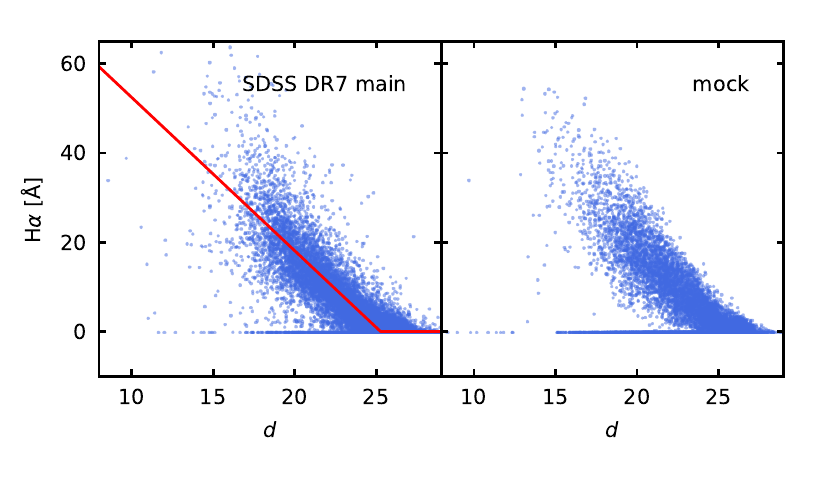}
\caption{Equivalent width of the H$\alpha$ line as a function of distance $d$ as defined by Eq.~\ref{eq:halpha} in the space of the first two continuum principal components. The red line shows our fit to the trend of SDSS data (left panel), whereas the right panel shows H$\alpha$ lines for the mock catalogue generated by our stochastic technique.}
\label{fig:halpha_fit}
\end{figure}

If a mock galaxy is selected as ``weak emission line + passive'' we generate the H$\alpha$ line only. The equivalent width of the line is calculated from the random continuum coefficients using Eq.~\ref{eq:halpha}. For each of the generated continuum coefficients we randomly determined whether any H$\alpha$ emission should be observable, then the emission line equivalent width baseline was determined based on Eq.~\ref{eq:halpha}, which was altered with a random value to match the observations (see Fig.~\ref{fig:halpha_fit}). At this step, taking the absolute magnitude into account was not necessary, as those galaxies, that exhibit only strong H$\alpha$ emission, are mostly quiescent star-forming galaxies, in which case the shape of the continuum correlates with the strength of the emission in a straightforward manner.

\subsection{Broadband colours and photometric error}
\label{sec:colors}

Once the stellar continuum was generated and emission lines were added, we calculated the SDSS broadband colours and apparent magnitudes using the standard formula of AB magnitudes and using the $r$ magnitude as a reference. Finally, random photometric noise was added to the apparent magnitudes based on the $1\sigma$ fits  introduced in Sec.~\ref{sec:photo_error}.

\section{Mock catalogue evaluation}
\label{sec:mock_eval}

We generated three mock catalogues: two mock catalogues that follow the parameter distributions of the SDSS main galaxy sample with a magnitude limit of $r =17.7$ which differed in sample size, and were used as training and test sample for the redshift estimations, and a faint sample with a magnitude limit of $r = 20.0$. Approximately 25.1\% of the galaxies in the two main sample catalogues have got strong emission lines, 37.7\% show $H\alpha$ emission only and 37.2\% are passive. These ratios in the faint galaxy sample are 26.9\%, 37.0\% and 36.1\%. The fainter sample consists of 30000 galaxies in total, while the main sample mock catalogues number 30000 and 24254 objects, with a similar relative number of galaxies from different classes. By simulating a separate catalogue for  both the training and test set of the photo-z evaluation, we ensured that the estimation precision will not be affected by the non-identical ratios of different galaxy classes between the two sets. The fainter sample consists of mock galaxies generated from the same luminosity functions but have got significantly larger simulated relative photometric error. The main parameters of the cosmology and luminosity function, as well as the applied magnitude cuts, are summarized in Tab.~\ref{tab:mockparams}. 

\begin{table}
\begin{tabular}{l | c | r | l }

parameter & symbol & value & unit \\
\hline
\hline

\multicolumn{3}{l}{Cosmology:} \\
\hline
matter density 			& $\Omega_M$			& 0.287 & \\
dark energy density		& $\Omega_\Lambda$	& 0.713 & \\
curvature				& $\Omega_k$			& 0  & \\
Hubble constant			& $H_0$				& 69.32 & km~s$^{-1}$~Mpc$^{-1}$\\

\hline
\multicolumn{3}{l}{Luminosity function:} \\
\hline
characteristic magnitude & $M_{r*}$ 			& -21.53 & mag \\
slope					& $\alpha$ 			& -1.23 &  \\

\hline
\multicolumn{3}{l}{Apparent magnitude limits:} \\
\hline
training set 			& $m_{r}$ 			& 17.7 & mag \\
faint sample				& $m_{r}$ 			& 20.0 & mag \\

\end{tabular}
\caption{Parameters of the mock catalog.}
\label{tab:mockparams}
\end{table}

To compare the main properties of one main sample mock catalogue to the SDSS main galaxy sample, we plot the normalized distribution of redshift, absolute and apparent magnitude in Fig.~\ref{fig:catalog_hists}. The distribution of the simulated absolute magnitudes matches the observational data well, whereas the redshift distribution, as expected, does not account for the effects of large scale structure, as these values were drawn from a uniform distribution according to the comoving volume. The distribution of the apparent magnitudes also matches that of the SDSS main sample.

As a verification of the emission line generation algorithm, we plot the BPT diagram of the mock catalogue in the right panel of Fig.~\ref{fig:bpt}. While it is not a perfect match to the SDSS sample, the AGN and star forming sequences are clearly distinguishable and show sufficient variance to cover the parameter space of the original data. However, the mock catalogue seems to miss the galaxies with very strong emission on both, the AGN and the star-forming side. As a result, the simulated BPT diagram appears to be shrank along the horizontal axis. This suggests, that the use of Gaussian Mixture Models is not sufficient to model the population of galaxies, as the outstandingly strong emission lines were not reproduced. The use of heavier tailed distributions could result in a complete reproduction of the set of emission lines, however, we chose to use the Gaussian model for this study.

The left panel of Fig.~\ref{fig:halpha_lines} shows the normalized histograms of H$\alpha$ equivalent widths for both the SDSS sample and the mock catalogue, whereas Fig.~\ref{fig:halpha_fit} shows the scatter plots of the equivalent width values as a function of distance $d$, as defined by Eq.~\ref{eq:halpha}. According to Fig.~\ref{fig:halpha_lines}, the mock catalogue reproduces the distribution of the main galaxy sample, although some minor discrepancies can be observed: our algorithm produces more galaxies that exhibit strong H$\alpha$ emission than the real galaxy sample would suggest, as visible on the left panel of Fig.~\ref{fig:halpha_lines} and the right panel of Fig.~\ref{fig:halpha_fit}. The most probable source of this discrepancy is the simplicity of the model that was used to describe the dependency between the equivalent widths and the distance parameter $d$ (Eq.~\ref{eq:halpha}). Since the first two principal component of galaxy continua define the colour of the galaxy and the low $d$ regime is connected to primarily bluer colours (as both principal component coefficients are low, which regime is primarily connected to strong emission line spirals, see Fig.~\ref{fig:cont_pcs}), we suspect that the algorithm produces more blue galaxies than what we observe in case of the real sample, which are then assigned to accordingly stronger equivalent width values by the simple model. However, as this discrepancy is negligible in terms of relative occurrence, we chose not to alter the linear model for the simulation process.

\begin{figure}
\includegraphics{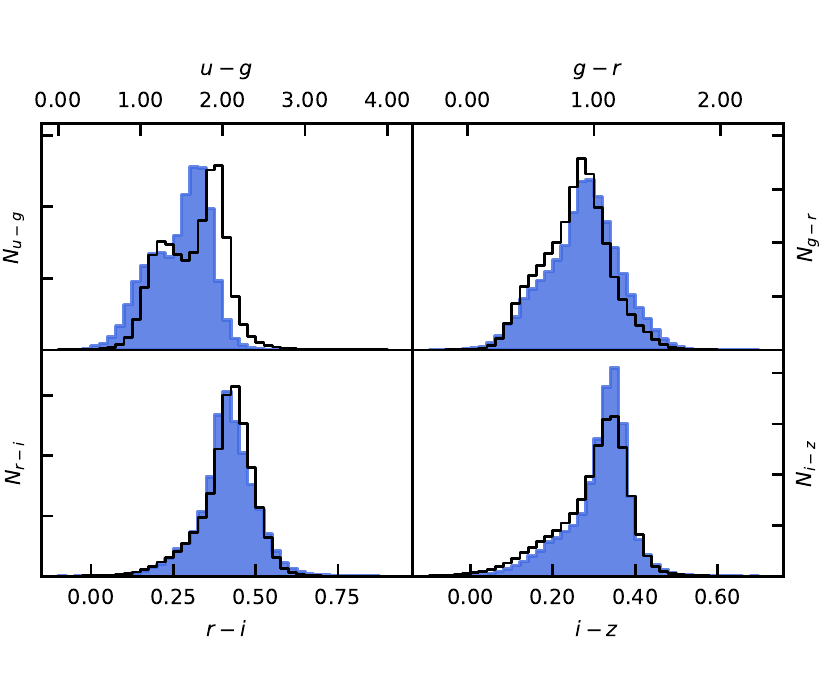}
\caption{Normalized distributions of broad-band colour indices for the SDSS emission line galaxy sample and the mock catalogue generated by our method. As it is visible from the similarity of the distributions, our model can results in realistic colours. For a discussion of the apparent offset in $u-g$, see the text.}
\label{fig:color_dist}
\end{figure}

Fig.~\ref{fig:color_dist} shows the broadband colour distributions of the mock catalogue and the real sample. Although the mock catalogue generation process does not involve any direct constraints on the broadband colours, the PCA-based procedure reliably reproduces the observable colour distributions with the exception of $u-g$. The offset in $u-g$ is attributed to the fact that observed spectra had to be extended in wavelength coverage towards to blue by the use of models in order to compute the SDSS $u$ broadband magnitude, as described earlier in Sec.~\ref{sec:cont_pca}. When fitting the continuum, the near UV end of the spectrum is not very well constrained by the longer wavelength flux. As we used the same mock catalogue generation algorithm to prepare the training sets and the test sets for photo-z evaluation, this bias is not expected to affect the results of the empirical algorithms. In case of template-based photo-z, which we don't investigate in this study, care must be taken and the bias need to be corrected or the simulated $u$-band has to be left out of the evaluation.

\begin{figure}
\includegraphics{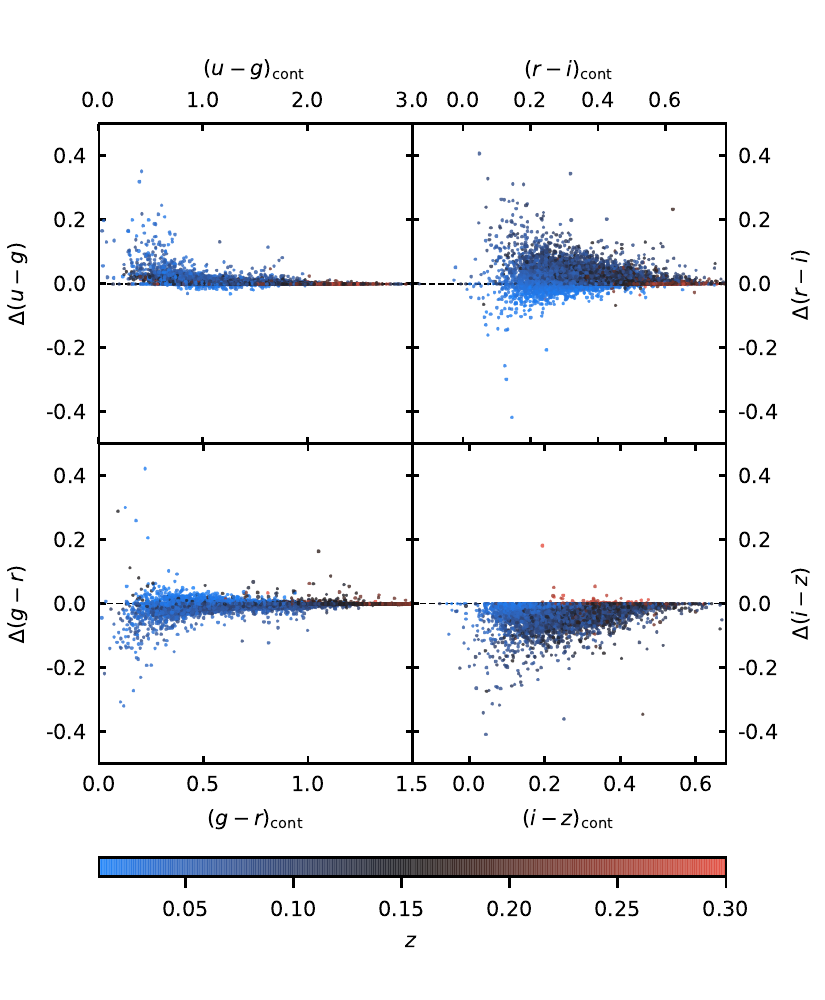}
\caption{The effect of emission lines on SDSS broad-band colour indices. Continuum-only colours are on the $x$-axis, the $y$-axis represents the difference caused by adding the emission lines. Colour coding is by redshift. Since we matched the flux limited SDSS DR7 spectroscopic sample, galaxies at higher redshifts tend to be passive LRGs with no significant emission lines.}
\label{fig:cc_lines}
\end{figure}

\begin{figure}
\includegraphics{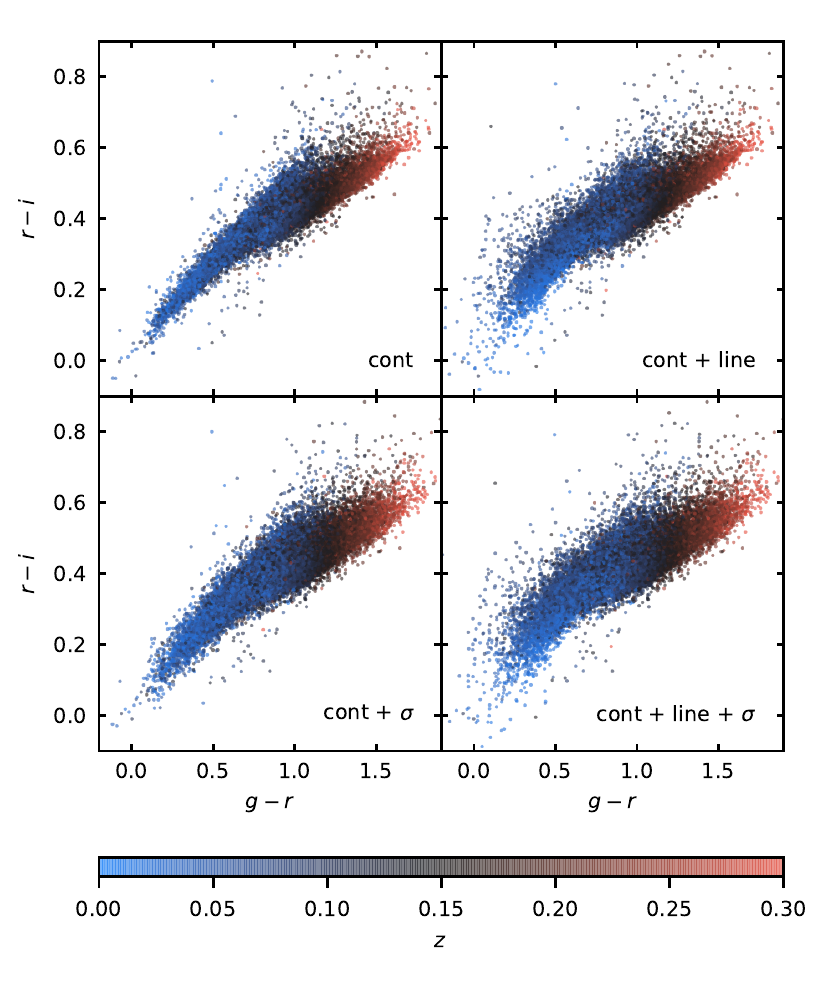}
\caption{The effect of emission lines and photometric error on SDSS broad-band colour indices, plotted in the $g-r$, $r-i$ colour-colour space for the bright $m_r < 17.77$ part of the mock catalogue. The panels show colours calculate with and without emission lines and photometric error. At this low level of photometric error, the broadening of the colour space coverage is primarily due to emission lines.}
\label{fig:cc_colors}
\end{figure}

Since the mock catalogue allows for turning emission lines on and off, we can visualize the effect of strong lines on broadband colours. In the left four panels of Fig.~\ref{fig:cc_lines}, we plot the SDSS broadband colour indices of the emission line mock galaxies against the colour indices of the pure stellar continua to emphasize the effect of strong emission lines on galaxy colours. As it is expected, the scatter in colours due to the emission lines affects low redshift blue galaxies more prominently since high redshift galaxies in the SDSS main galaxy sample are LRGs due to the shallow flux limit. The right four panels of Fig.~\ref{fig:cc_colors} show the colour distributions of the main sample mock galaxies with and without emission lines and photometric error. According to the plot, at these bright magnitudes, the main source of scatter in colours is the presence of emission lines and not photometric noise. Hence, the main source of error in template based photometric redshift estimation methods is very likely the limited variance of emission lines of the templates. On the other hand, as we will show later, the scatter due to emission lines help empirical photometric redshift estimation methods significantly at low redshifts.

\section{Photo-$\lowercase{z}$ algorithms}
\label{sec:methods}

In this section we briefly describe the algorithms we tested with mock catalogues consisting of weak and strong emission line galaxies. The algorithms chosen are intentionally simple compared to the state-of-the-art empirical photo-$z$ methods as our primary intention is to assess the effect of emission lines on photometric redshift estimation and not to optimize the algorithms themselves. Both tested methods are empirical and require a training set with known photometry and redshift. Despite their relative simplicity, the techniques perform well and can be used to yield error estimates on photometric redshifts as well.

\subsection{The $k$-nearest neighbour method with local linear regression (LLR)}

The simplest empirical methods work by finding galaxies in the training set with very similar colours to galaxies with unknown redshifts. Similarity is very often defined as a Euclidean distance in the D-dimensional space of broadband colour indices or magnitudes, a choice lacking any physical meaning. The $k$-nearest neighbour ($k$NN) method, in particular, works by finding $k$ galaxies in the training set that are the closest to a given galaxy with no known redshift. In the simplest case the photometric redshift is taken to be the (weighted) average of the known redshifts of the nearest neighbours. Computationally efficient algorithms are available in the \texttt{scikit-learn} Python package to speed up neighbour look ups \citep{scikit-learn}. 

\cite{Beck2016b} further developed the $k$NN estimator method to mitigate the effect of outliers in the training set and compute a realistic error on photo-$z$, as well as to flag photometric galaxies which require extrapolation from the training set. The algorithm starts by finding the $k$ nearest neighbours of a photometric galaxy in the training set, but instead of simply averaging the known redshifts, it performs a linear fit of redshift as a function of a data vector $x_{i,l}$ defined from magnitudes and colour indices, where $i = 1, 2, ..., k$ indexes the nearest neighbours. The data vector of the galaxy with unknown redshift is denoted by $x_{0, l}$. The redshift is modelled in the form of
\begin{equation}
	z_\textrm{phot}(x_{i,l}) = c + a_l x_l,
	\label{eq:knnphotoz}
\end{equation}
where $c$ is a constant offset and $a_l$ are linear coefficients. The values of $c$ and $a_l$ are found by minimizing the expression
\begin{equation}
	\chi^2 = \sum_{i=1}^{k} \frac{ \left( z_i - c - \sum_l a_l x_{i,l} \right)^2}{w_i},
\end{equation}
hence the name of the method (local linear regression, LLR). The weight $w_i$ can take the photometric error of training set galaxies and the distance of the nearest neighbours from $x_{0,l}$ into account. The redshift of the photometric galaxy is found by substituting $x_{0,l}$ into Eq.~\ref{eq:knnphotoz}.

The goodness of the linear fit over the $k$ nearest neighbours, which can be considered as an approximation to the error of the photometric redshift, is calculated by \cite{Beck2016b} as
\begin{equation}
	\sigma_z = \sqrt{ \frac{\sum_{i = 1}^{k} \left( z_i - c - \sum_l a_l d_{i,l} \right)^2 }{k} }.
\end{equation}
This quantity can also be used to constrain items of the training set used for interpolating the redshift from colours and magnitudes. Once the $a_l$ coefficients are determined, one can substitute the $x_l$ data vectors into $z_\textrm{phot}(x_{i,l})$ and verify if $z_\textrm{phot}(x_{i,l}) < 3 \sigma_z$ holds. Nearest neighbours not satisfying the constraint can be removed and the fitting repeated.

When interpolating a multivariate function $f$ at $x_0$ from the nearest neighbour known values at $\left\{ x_i \right\} = NN(x_0, k)$, one can define \textit{extrapolation} as the case when $x_0$ is outside the convex hull of $\left\{ x_i \right\}$. The volume of the convex hull can also be used to define the quality of interpolation.

\subsection{The Random Forest method}

Random Forest (RF) is another empirical method applicable to photometric redshifts that also yields reliable error estimates \citep{Ho1995}. The method works by building a large ensemble of decision trees over randomized subsets of the training set. Bootstrapping is done by selecting random subsets of training set galaxies, as well as selecting random subspaces of the training set data vectors, i.e. magnitudes and colour indices. Randomizing the subspaces used to build the decision trees help avoid overfitting. Best parameter estimates are taken by averaging the predictions from the individual trees over the ensemble while the standard deviation of predictions gives realistic error. A significant advantage of random forests over $k$NN estimates is that RFs do not require a metric over the colour space, hence do not involve any artificial, non-physical quantities. We used the RF implementation of the \texttt{scikit-learn} Python package \citep{scikit-learn}.

\subsection{Self-organizing maps (SOM)}

Self-organizing maps \citep{Kohonen1982} are a simple form of artificial neural networks that allow competitive learning. SOMs can map high dimensional data vectors onto a few, typically two-dimensional grid. These maps tend to preserve some topological properties of the original data. For instance, if galaxies with similar redshifts are located near each other by some metric in the original space of magnitudes or colour indices, they tend to be mapped near each other by SOM as well. While SOMs have got also been successfully used for photometric redshift estimation \citep{CarrascoKindBrunner2014}, in the present paper we focus on the visualization capabilities of the method.

To construct our SOM, we consider the data set as data vectors $x_{i, m}$, where $i = 1, ... N$ indexes the $N$ galaxies of the sample and $m = 1, ... M$ indexes the $M$ broadband colours. The $x_{i, m}$ vectors are usually called \textit{features} in the computer science literature. The SOM is defined as a regular grid of $K \times K$ neurons, also referred to as \textit{cells}, that are organized in a 2D lattice pattern. Each of the $K \times K$ neurons has $M$ weights which we will denote by $w_{k, m}$, where $k = 1, ... K \times K$ indexes the neurons. The weight vectors will be initialized to random values which will be iteratively updated during the training process. The training process will consist of $T$ passes, or \textit{training epochs}, over the data set, each time in a randomized order. At each iteration, the algorithm takes a galaxy and finds the SOM cell with index $k_\mathrm{best}$ which has the weight vector $w_{k, m}$ closest to feature vector $x_{i, m}$ of the galaxy in least squares sense, i.e.
\begin{equation}
k_{\textrm{best}} = \argmin_k \sum_{m=1}^M \left[ x_{i, m} - w_{k,m}^{(t)} \right]^2,
\end{equation}
where the $t = 1, ... T \times N$ superscript denotes the value of the weights at epoch $t$. Once $k_{\textrm{best}}$ is found, the weights of all SOM cells are updated by the rule following \cite{CarrascoKindBrunner2014}. Updates to the weights are scaled according a neighbourhood function which can take into account the SOM cells' distance from the cell $k_{\textrm{best}}$ by the Euclidean metric of the 2D embedding space of the neurons. Updates are calculated as
\begin{equation}
w_{k, m}^{(t+1)} = w_{k,m}^{(t)} + \alpha^{(t)} \cdot H_{\mathrm{best},k}^{(t)} \cdot \left[ x_{i, m} - w_{k, m}^{(t)} \right],
\end{equation}
where $\alpha^{(t)}$ is the learning rate and $H_{\mathrm{best},k}^{(t)}$ is the neighbourhood function. The index $i$ denotes the feature vector that we use to calculate the updates in iteration $t$.

We used a decreasing learning rate in the form of
\begin{equation}
\alpha(t) = \alpha_s \left(\frac{\alpha _e}{\alpha _s}\right)^{t/(T \cdot N)},
\end{equation}
where $\alpha _s$ and $\alpha _e$ are the starting and ending learning rates. We used the values $\alpha_s = 1$ and $\alpha_e = 0.1$.

For the purposes of the neighbourhood function we used a Gaussian kernel in the form of
\begin{equation} 
H_{\textrm{best}, k} = \mathrm{e}^{-D_{\mathrm{best},k}/(\sigma^{(t)})^2}
\end{equation}
where $D_{\mathrm{best},k}$ is the Euclidean distance between SOM cells $k_\mathrm{best}$ and $k$, as measured in the embedding space of the SOM neurons. The $\sigma^{(t)}$ parameter is the width of the neighbourhood function which decreases according to the formula
\begin{equation}
\sigma^{(t)} = \sigma_s \left(\frac{\sigma_e}{\sigma_s}\right)^{t/(T \cdot N)}
\end{equation}
where $\sigma_s$ and $\sigma_e$ denote an initial and final values of the kernel width. The former is usually set to the size of the map, while the latter should be roughly the size of a cell, which corresponds to $\sigma_s = K$ and $\sigma_e = 1$ if the cells located at unit distance from each other in the lattice.

Once the SOM is constructed, after the last iteration we can order each galaxy to one of the cells of the map by finding the cell $k_\mathrm{best}$ with weights closest to the vector of broadband colours. We chose the values $K = 58$ and $T = 300$, whereas the number of galaxies was approximately $N \approx 30{,}000$. Since we created the SOMs solely for visualization purposes, we wanted to create as big maps as possible while avoiding overfitting which occurs when the number of cells exceeds the number of input data vectors. By setting $k$ to 58, we ensured that structures in the map will appear, yet the average number of galaxies per cell will still be high enough for these structures to be statistically meaningful.

\begin{figure*}
\includegraphics[trim={1.5cm 1cm 0 0},clip]{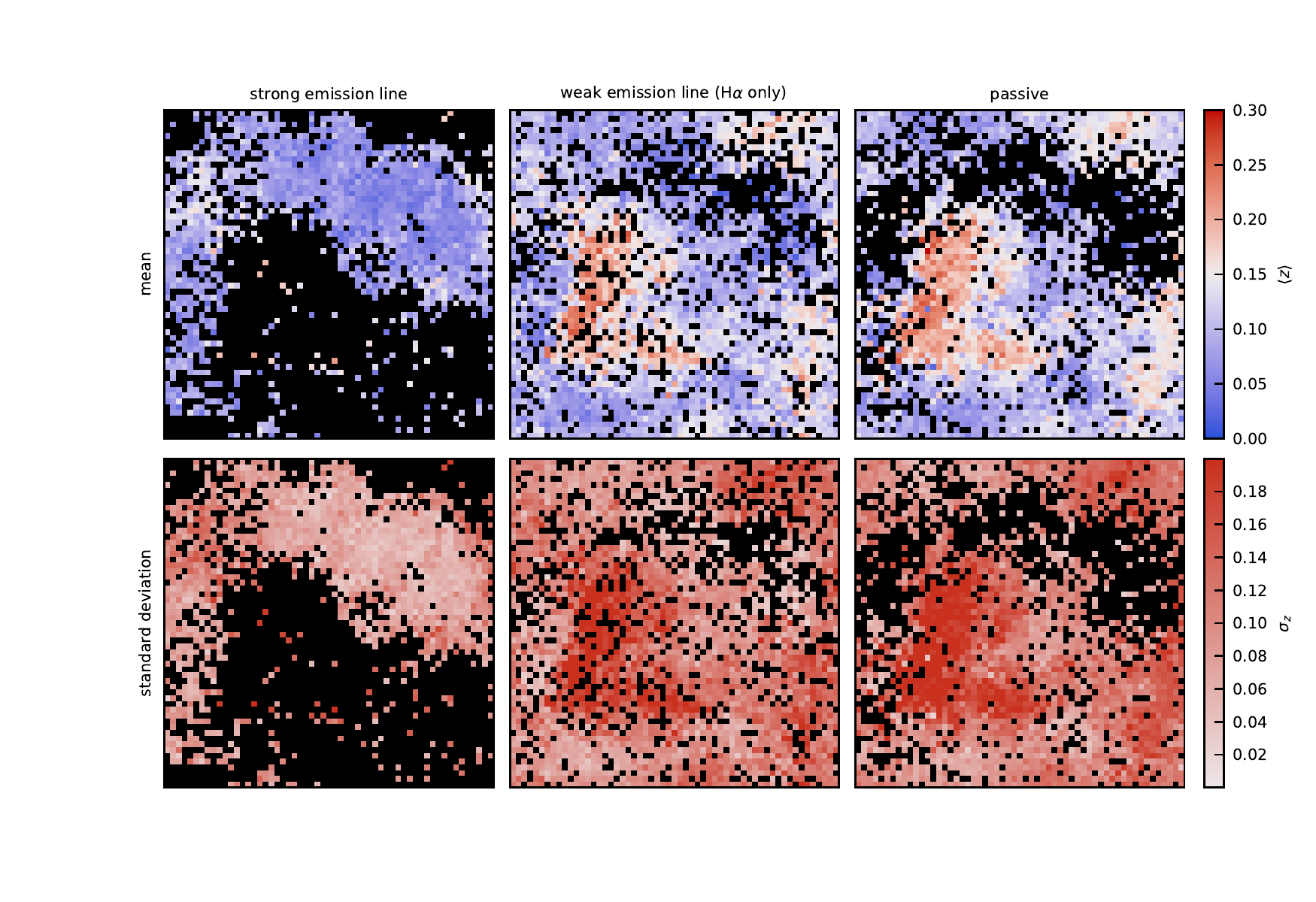}
\caption{Self-organizing maps generated from the mock catalogue using synthetic broad-band colours as input. Panels in the top row show the mean redshift $\langle z \rangle$ in every SOM cell for the strong emission line, weak emission line and passive subsample, respectively. The bottom row shows the $\sigma_z$ standard deviation of redshift in each SOM cell. Black SOM cells contain zero or one galaxy of the particular type. Strong emission line galaxies fill in significant gaps in the broad-band colour space, hence are necessary for valid photometric redshift estimation. Note that the axes are not physical, but are merely indices into the map.}
\label{fig:som}
\end{figure*}

Once the whole catalogue was distributed among the SOM cells, we calculated the moments of the redshift distribution of galaxies in each cell, then coloured the SOM coordinate bins in accordance with these values. The resulting SOMs of the bright sample are plotted in Fig.~\ref{fig:som}. The upper row of images show the map of projected broadband colour values as cells coloured by the average redshift of the galaxies belonging to that bin, whereas the lower row shows the same map coloured by the standard deviation of the galaxy redshifts in each bin. The different columns of the figure show the map filtered for the three types of galaxies. All three maps show patches that are devoid of galaxies, which shows that none of the three considered emission line types can cover the entire colour-field spun by the complete mock catalogue. More importantly, the stacked map of the weak emission line and passive galaxies cannot cover the entire field either, which implies the importance of inclusion of strong emission line galaxies for template based photometric redshift evaluation methods for yielding statistically correct results. If these galaxies were missing from the training sample, then the redshift of some galaxies would be determined through extrapolation, which would lead to biased results.

\begin{figure*}
\includegraphics[trim={0 0 0 0},clip]{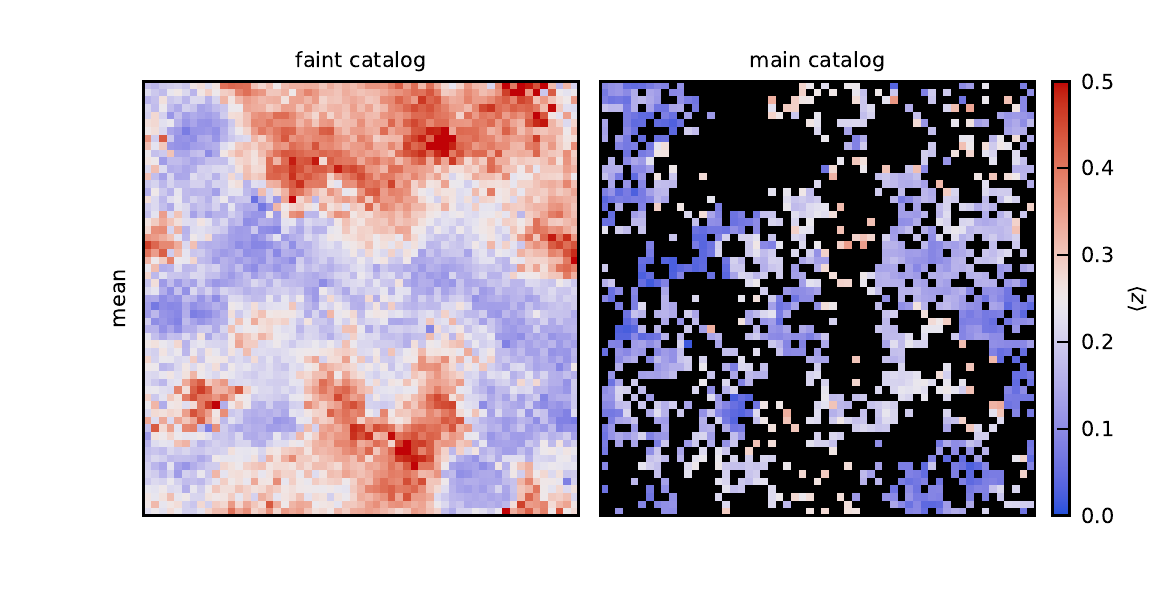}
\caption{Self-organizing maps generated from the simulated mock catalogues using synthetic broad-band colours as input. The left-hand side panel shows the self-organizing map of the faint mock catalogue, while the right-hand side panel shows the main sample mock catalogue projected to the map of the faint sample. The large amount of black squares on the latter map indicate the absence of galaxies with certain broadband colours.}
\label{fig:faintsom}
\end{figure*}

A common problem of empirical photometric redshift estimations is that the reference catalogue with measured spectroscopic redshift values is usually brighter in terms of limiting magnitude than the test set, which can result is significant differences between the distribution of broadband colours of the training and test sets and can ultimately influence the precision of the photometric redshift estimations. Self-organising maps are also useful to assess how representative a training set is for empirical photo-z. For example, redshifted faint blue galaxies can occupy the same colour locus as low redshift red galaxies. Fig.~\ref{fig:faintsom} shows the self-organising map generated from the fainter mock catalogue (with a magnitude limit of $r < 20$) and the brighter mock catalogue projected to the calculated map. The large amount of empty pixels on the brighter catalogue map suggests that the fainter catalogue covers colour ranges that are not present in the brighter catalogue. This indicates that there exists a subset of galaxies in the fainter sample for which an empirical photo-z estimate based on a brighter training set will be biased. 

\section{Photo-$\lowercase{z}$ evaluation}
\label{sec:results}

To assess the influence of the presence of emission lines on the performance of the various photo-z estimators, we investigated each of the three methods (Sec.~\ref{sec:methods}) on the mock catalogues in four different settings:
\begin{enumerate} 
\item{continuum only, where the emission lines were excluded from the spectra and the synthetic magnitudes did not contain the photometric noise term}
\item{continuum+photometric error, where synthetic magnitudes were calculated from the continuum and photometric error was added}
\item{continuum+emission lines, which represents the case of perfect measurements, as the simulated photometric error is not added to the magnitude values}
\item{continuum+emission lines+photometric error, which represents the realistic case.}
\end{enumerate}
For each setting, the number of neighbours for the kNN and LLR estimators was set to 30. The training set was chosen to be the smaller brighter mock catalogue for each of the estimations. It is important to note, that for each of the four settings outlined above the spectral sets with the same characteristics were used for both the training and the test set, e.g. for the pure continuum case, not only the test magnitudes were calculated with the exclusion of emission lines and the photometric error, but the training set magnitudes as well.
 
\begin{figure*}
\includegraphics[trim={0 1.5cm 0 0},clip]{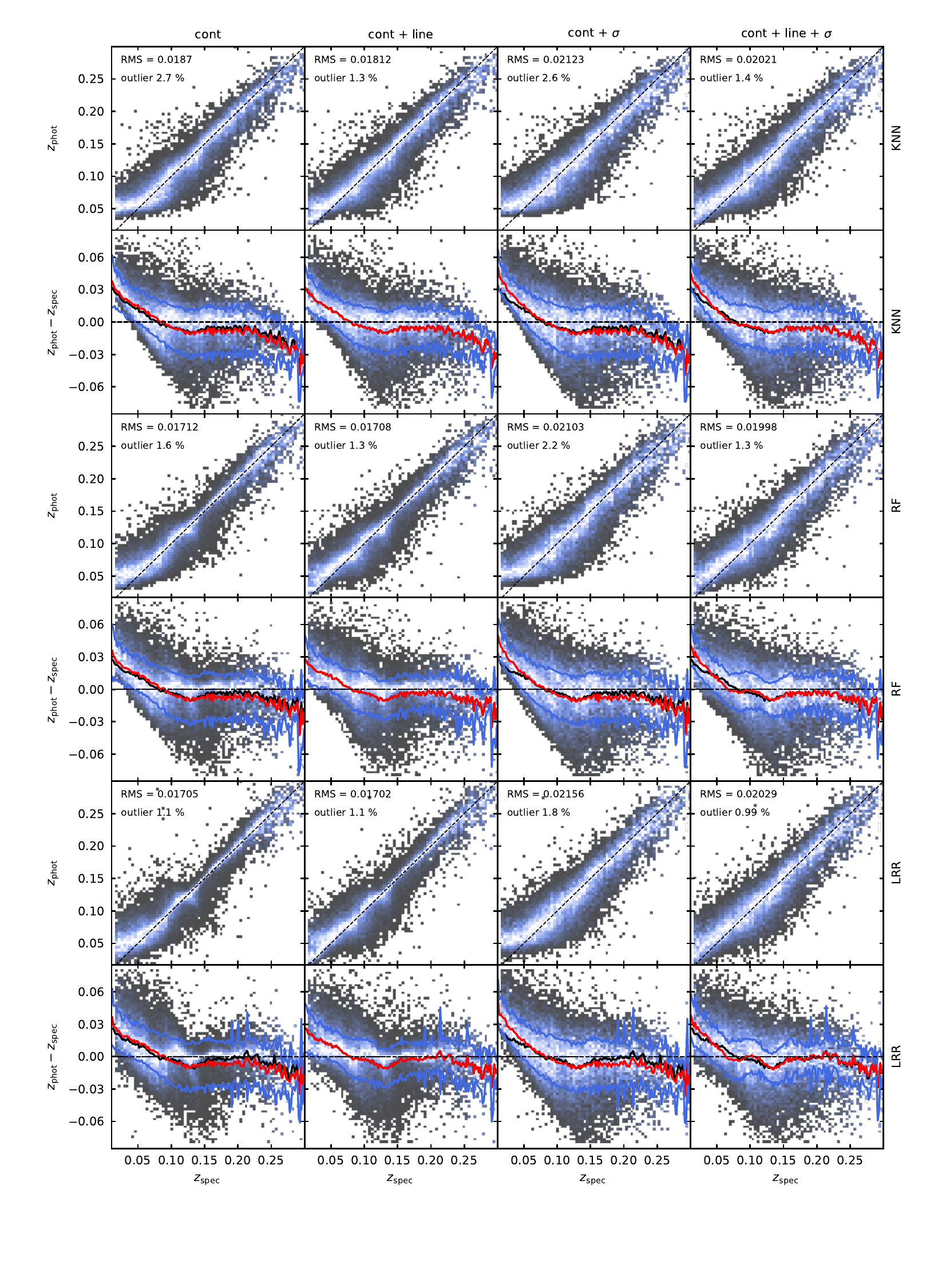}
\caption{Evaluation of the three photo-$z$ methods for the bright mock catalogue using the $k$-NN, random forest and local linear regression methods with and without emission lines and photometric error. Odd rows show the photometric redshift estimate $z_\mathrm{phot}$ as a function of the model redshift $z_\mathrm{spec}$. Even rows show the $\Delta z = z_\mathrm{phot} - z_\mathrm{spec}$ bias as a function of model redshift. The red curve highlights the mean value of the bias, whereas blue curves represent $1\sigma$ intervals. The solid black curve is the bias of the \texttt{cont + line} case plotted for reference in each column. Outlier per cents are quoted for $3\sigma$ outlier counts. See text for discussion.}
\label{fig:pz_bright}
\end{figure*}

Fig.~\ref{fig:pz_bright} shows the performance of the photo-z estimators in the various cases. The odd rows show the estimated photometric redshift, while the even rows show the bias as a function of the simulated redshift. For every estimation the RMS error is expressed in the form
\begin{equation}
\Delta_{\textrm{RMS}} = \sum_i \frac{\Delta z_i}{1 + z_i}.
\end{equation}
For each of the estimations we also calculated the outlier rate by flagging each of the galaxies an outlier for which the difference between the simulated and estimated redshifts differed with more than $3\sigma$. It has to be noted, that these galaxies were still included in the calculation of the RMS, since in case of realistic estimations, we cannot detect outliers on this basis.

\begin{figure}
\includegraphics{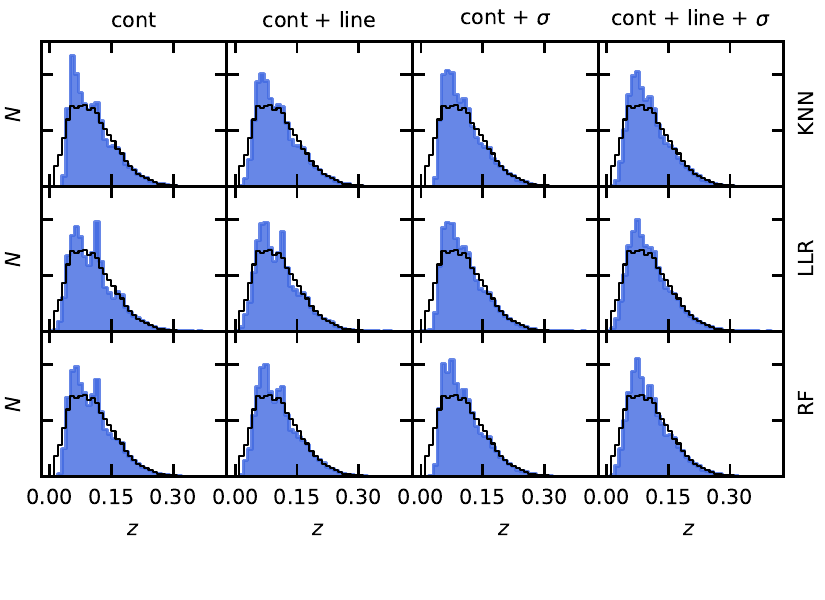}
\caption{The photometric redshift distributions obtained with the different estimators (blue) compared to the redshift distribution of the mock catalogue (black).}
\label{fig:pz_Nzdz}
\end{figure}

The first characteristic that can be observed for each of the estimation methods is the degrading effect of the photometric error, which reduces the accuracy of the estimates significantly. The second feature, which is also inherent in all the various methods used is the effect of emission lines; once the emission lines are superimposed on the continuum spectra, the quality of the estimation improves compared to the purely continuum based case. This effect could also be observed in the case extended with the photometric errors. In terms of statistics, the lines have moderate effect compared to the photometric precision of SDSS, as the RMS in case of included measurement errors decreases with only few percent due to the addition of lines. The effect is more distinguishable in the bias plots in Fig.~\ref{fig:pz_bright}: when the emission lines are included, the absolute bias of the estimation decreases for all redshift values, but most significantly for the lowest redshift galaxies. This effect can be explained by the redshifting of emission lines across different wavelength which resolve degeneracies present in the colour space of the continua and alter the position of the galaxy on the colour-colour diagram non-negligibly compared to the purely continuum based case. Accordingly, the emission lines reduce the possibility of having objects of different redshifts close to each other in the colour space, thus it improves the accuracy of the photometric redshift estimations. This effect is mostly significant only for the lower redshift galaxies due to the limitations of the SDSS sample itself, but more recent spectroscopic surveys are able to detect fainter strong emission line galaxies, which can allow for the investigation of the degeneracy resolution on much higher redshifts as well. The resulting photometric redshift distributions are shown in Fig.~\ref{fig:pz_Nzdz}.

Some differences between the bias plots of different settings can also be explained by the addition of emission lines. The improvement of bias in the low redshift regime ($z < 0.05$) can be explained by the filter shift of the H$\alpha$ line, which adds a flux excess to the $i$ filter and thus it partly resolves the degeneracies in this redshift range. The second observable difference that can be accounted to the presence of emission lines is observable in the $z>0.13$ regime: in this range the absolute value of the bias is decreased which can accounted to the filter shift of the [OIII] and H$\beta$ emission lines. Thus, the emission lines add sharp features to the spectra of galaxies, which will result in a more precise estimation than what we observe in case of purely continuum based estimations.

\begin{figure*}
\includegraphics{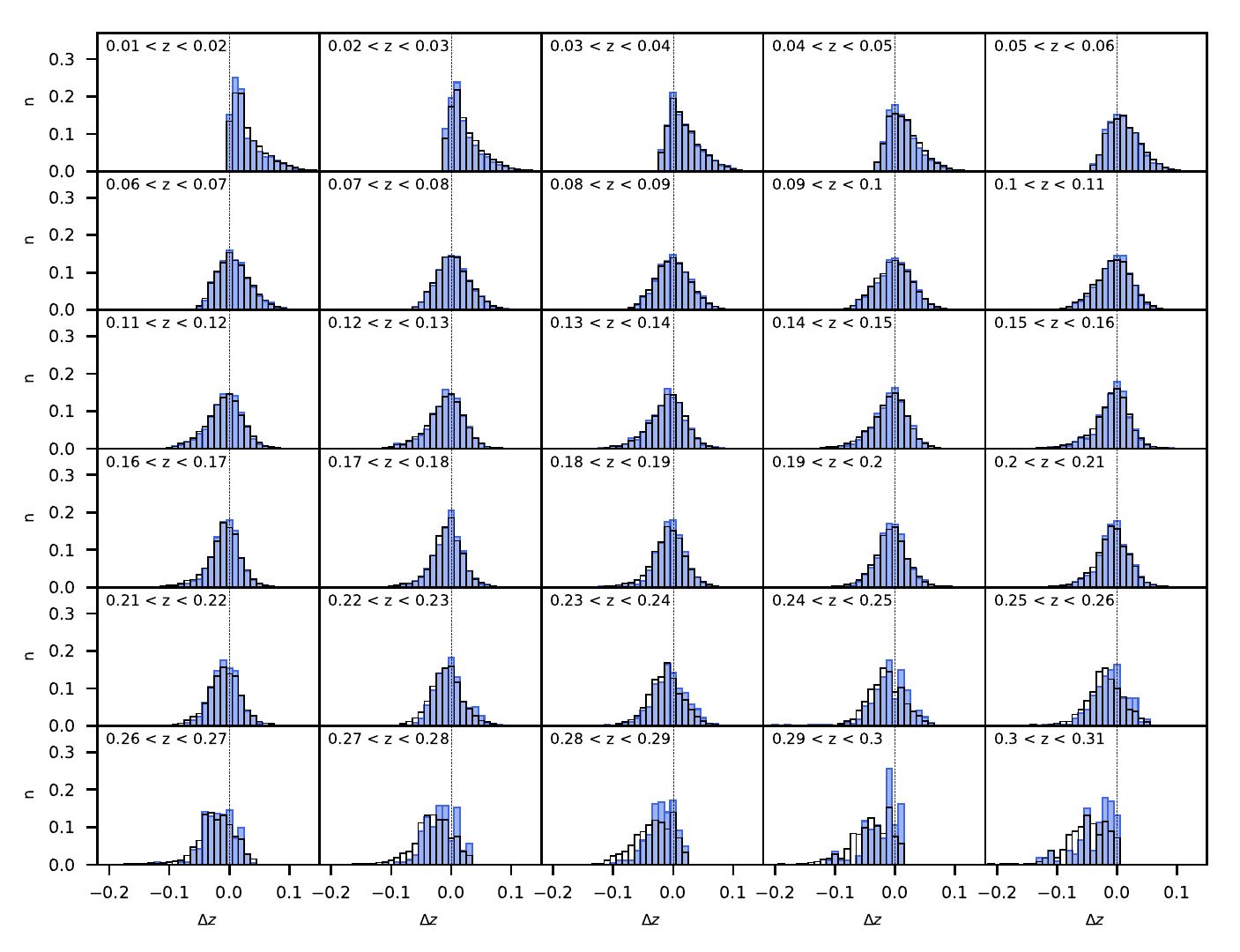}
\caption{The normalized redshift distributions of the neighbours of galaxies at the various redshift ranges. Each galaxy has been sorted into one of the ranges and the horizontal axis of the figures show the differences from the central redshift of the corresponding range. At lower redshift ranges the neighbouring galaxies have higher average redshift than the central redshift value, which implies that simplest average based estimators overestimate the redshift of the objects, due to the lack of negative redshift objects. The case for the higher redshift ranges is the opposite, where systematic underestimation can be observed, due to the finiteness of the catalogue. In the intermediate redshift range ($0.05<z<0.13$) the histograms resemble normal distributions that are centred on the central redshift values of the ranges, which implies that the estimation bias should disappear or minimize at this range.}
\label{fig:tsHists}
\end{figure*}

The presence of a bias in each of the estimation methods suggests that it is independent of the estimator, and it originates from the training set, thus it can be corrected for. According to our hypothesis, the bias of the estimations is determined primarily by how the redshifts of the neighbour galaxies are distributed on average. Due to degeneracies and various errors, galaxies from different redshifts are mixed in the colour space, where the scale of the mixing is significantly affected by how the galaxies are distributed according to redshift: if a redshift range contains a large amount of galaxies, then the probability of colour space scatter is higher, i.e. it is more probable to find a galaxy of this redshift with colour values that should be exhibited by galaxies on either lower or higher redshifts. Such colour space scatter causes the neighbour finding algorithms to systematically pick up galaxies that are farther away in terms of redshift, which ultimately causes the observed bias. To test our hypothesis, we calculated the redshift distributions of neighbours for the brighter mock catalogue in various redshift ranges: we calculated the redshift histograms for every galaxy in each redshift range, then added these together, to obtain a single average distribution in each redshift range, which are shown in Fig.~\ref{fig:tsHists}. In the central ($0.06 < z < 0.2$) redshift range the distributions are mostly centred on the correct redshift values, thus in this range the estimation yields results that are unbiased. However, on the lower and higher redshift ranges these distributions are systematically shifted toward one side, which causes the observed bias: since the estimators applied by us are primarily sensitive to the average of these distributions (especially the $k$NN and RF methods), which in this case do not characterize the distributions properly, they return biased estimation values for such systematically skewed distributions. Although the LLR method is more robust against outliers by its construction (Sec.~\ref{sec:methods}), the skewness of these distributions also cause similar biases than in the case of $k$NN and RF. To correct the bias of the estimations, non-average based (for example, mode based) algorithms would be required, that yield correct results even for skewed redshift distributions (however, their statistics would then be inevitable more complex). In accordance with the calculated redshift distribution of colour-space neighbours, the problem arises from the limited redshift coverage of the mock catalogue: in the lower redshift range only galaxies from higher redshift can scatter into the colour neighbourhood, as there are no objects with negative redshifts. This causes the empirical methods to overestimate the redshift of the object in question, which ultimately leads to the observed bias. This also suggests that the bias at this range is determined by the luminosity function in first order, as it only depends on the spectral types and brightness of the simulated galaxies. On the higher redshift range the situation is quite the opposite, as only the galaxies from the lower redshifts can scatter among the neighbours, which leads to a systematic underestimation of redshift. This problem can be easily mitigated by extending the training set to higher redshifts, however this would leave the bias observed on lower redshift ranges unaffected.

\begin{figure*}
\includegraphics{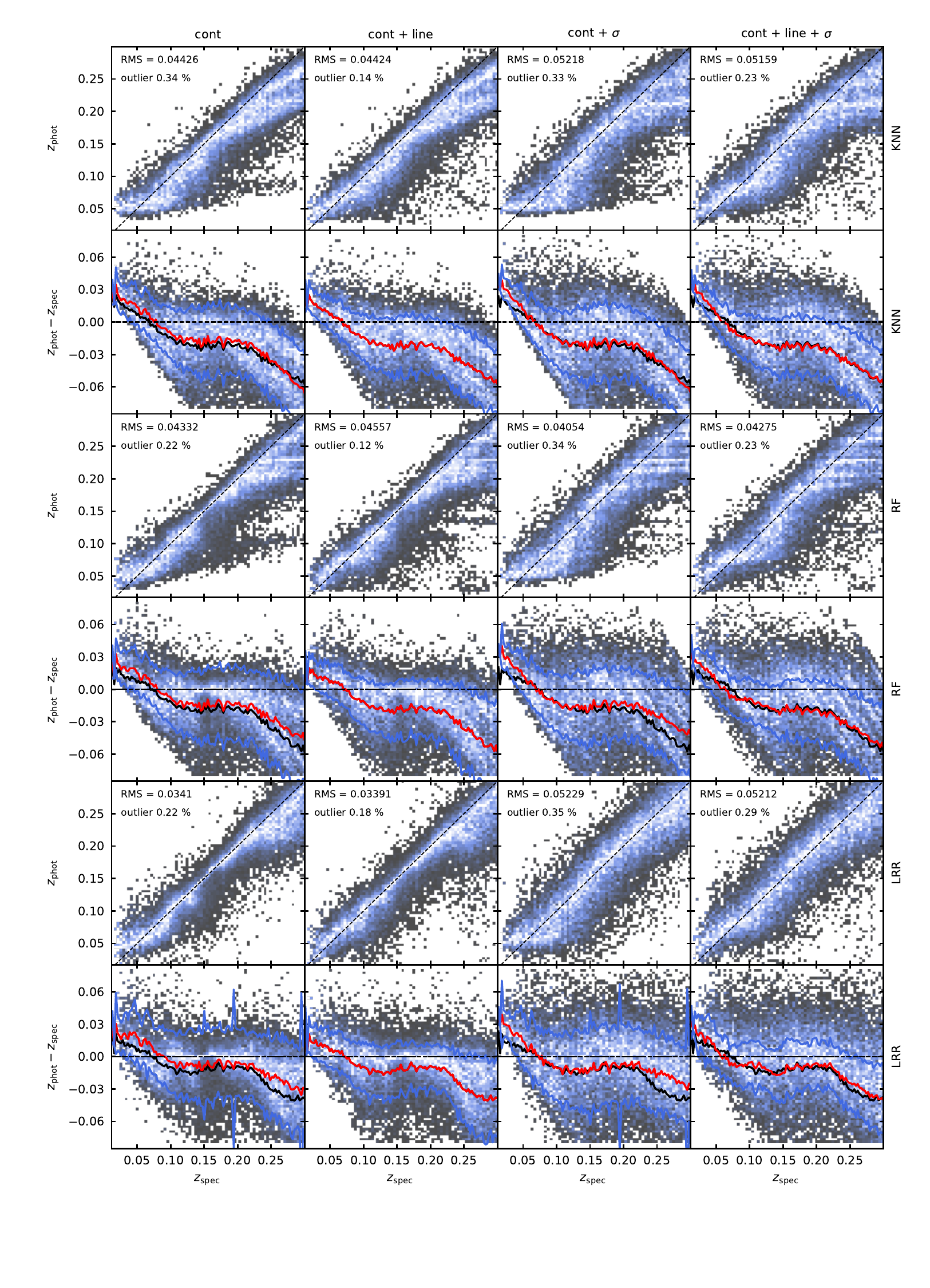}
\caption{Evaluation of the three photo-z methods with one main sample mock catalogue used as the training set and the faint mock catalogue as test set. The image layout is identical to the Fig.~\ref{fig:pz_bright}.}
\label{fig:pz_faint}
\end{figure*}

To test how well the brighter catalogues can be used for a photometric redshift estimation of fainter catalogues, we applied the described estimators on the brighter and fainter mock catalogues. For the training process, the same bright mock catalogue was used as in the brighter sample estimations described earlier in this section. The results of the photo-z estimations are plotted in Fig.~\ref{fig:pz_faint}. Since the limiting magnitude of the training and test sets are considerably different, the estimations are a lot less precise than in the case of the main sample mock catalogues. The largest fraction of this difference is caused mainly by the intermediate redshift faint blue galaxies, which are not present in the main sample mock catalogue. These fainter galaxies occupy an area on the colour-colour diagram that overlaps with the intermediate-high redshift part of the main sample diagram, which causes an extra uncertainty for the estimator. This causes a general underestimation of redshifts and an even more significant bias on larger redshifts ($z>0.25$). Another major difference between the two sets of mock catalogues is the significance of the photometric error: as the test set comprises fainter galaxies, the error terms on the photometric magnitudes will be greater, which causes a much greater scatter on the colour-colour diagram compared to the main sample case, which leads to greater uncertainties. In overall, the emission lines seem to produce the opposite effect compared to the main sample case, as in their presence we observe the bias to be shifted towards more negative values, which either causes the precision of the estimation to drop compared to the continuum based case (for RF) or either improves the statistical quality of the estimation (for kNN and LLR). This shift causes a less significant bias on the lower redshifts, which suggests that the emission lines resolve degeneracies at this range, similarly, to the main sample case. However, on larger redshifts they cause the estimation to become even less accurate by mixing the colour field. Notably, the outlier rate decreased for each of the methods when the emission lines were added.

For each of the methods a breakdown in precision can be observed around $z\approx 0.25$, which correspond to the limited magnitude and redshift range of the training set: as there is only a limited number of galaxies above this redshift value in the bright mock catalogue, which are mostly LRGs, the faint blue galaxies from the test sample will have neighbours from much lower redshifts, which leads to systematic underestimation of redshift and to an increased bias. Due to the large difference between the limiting magnitudes the less complicated methods, like the kNN and RF, perform less accurately, while the LLR, which allows for an outlier detection and $\sigma$ clipping during the estimation, can still yield relatively unbiased results. This well demonstrates the necessity of the more complex methods for correct estimations in case extrapolation capabilities of the training set to the test set are limited.

\section{Conclusions}
\label{sec:conclusions}

In this study we developed a method that can be used to generate a mock catalogue of galaxy spectra with realistic line strengths and distributions. To establish such a technique, we relied on semi-empirical methods: we calculated a basis for both of the continuum spectra and the set of emission lines and modelled the catalogue distributions using PCA and Gaussian Mixture Models. The mock catalogues, which consist of passive (no emission lines, or H$\alpha$ only) and active (complete set of emission lines) galaxies mimic the realistic broad band colour and line strength distributions. Using self-organizing maps we showed that the inclusion of galaxies with a complete set of emission lines is necessary to cover the entire colour space spanned by the real galaxy catalogues. By applying three different estimators on the various mock catalogues we examined how the presence of emission lines affect the precision of photometric redshifts. According to our results, the emission lines, which correspond to sharp features present in the spectra, can improve the estimation accuracy by resolving degeneracies in the colour-colour field. This effect was only observed significantly when the limiting magnitude of the training and test set were identical, otherwise the effects caused by the emission lines are blurred by the photometric uncertainties. Our simulated results on photometric redshifts showed a non-negligible estimation bias, which showed a clear trend with the redshift for each of the estimators. Our analysis showed that this bias is inherent in the training set itself, but could be corrected using non-average based empirical algorithms, which can take the skewness of the colour space neighbour redshift distributions into account as well.

The SDSS galaxy sample, which we attempted to match with our mock catalogue, contains significantly more blue galaxies that show strong emission lines at low redshift than at redshift higher than $z > 0.15$ where the sample is dominated by passive Luminous Red Galaxies. Since red galaxies occupy a well-known sequence in the SDSS colour-colour space, their photometric redshift estimates are dominated by the photometric error. On the other hand, at low redshift, due to the mixing of galaxies of different types, the colour-colour space can be highly degenerate. The presence of strong emission lines further complicates the scheme. According to the results from the Random Forest estimator, the effect of emission lines is comparable to that of photometric error, at least at the quality of SDSS spectroscopic observations ($m_r$ < 17.77). Uncertainties of photometric redshifts of future large mirror surveys with lower photometric noise will likely be dominated by the variance in strong emission lines also at greater depths / higher redshifts. Consequently, as an immediate next step to this study, we will investigate the extrapolation capabilities of the method outside the brightness interval of the typical spectroscopic training sets.

Template-based photometric redshift estimation techniques usually employ a small, carefully selected set of templates to fit broadband colours, hence are more prone to uncertainties caused by the emission lines. Through the described method one can create a realistic set of galaxy spectra with infinite signal-to-noise ratio, which could be used as a set of galaxy templates for further research. Since K-correction and physical parameter estimation requires template fitting, a careful investigation of the biases due to strong emission lines and limited template sets will be performed in the near future.

\section*{Acknowledgements}

This project was supported by the Hungarian grant 
NKFIH OTKA NN 129148. This project received support from the Hungarian Artificial Intelligence National Laboratory. The research work of LD is supported by the Schmidt Family Foundation.

\section*{Data availability}
The catalogue data underlying this article are available at \url{https://www.sciserver.org} and can be accessed through CasJobs, whereas the spectra are available at the SDSS Science Archive Server (SAS, \url{https://dr16.sdss.org/optical/spectrum/search}). The analysis files and the code can be accessed at the GitHub page of the author: \url{https://github.com/Csogeza/EmiPhotoZ}.




\bibliographystyle{mnras}
\bibliography{linephotoz}

\bsp	
\label{lastpage}
\end{document}